%% file: main.tex
\documentclass[sigconf]{acmart}
\usepackage{color}

\usepackage{graphicx}
\usepackage{subfigure} 
\usepackage{float}

\usepackage{booktabs}
\usepackage{multirow}
\usepackage{color, colortbl}
\definecolor{LightCyan}{rgb}{0.88,0.88,0.88}

\usepackage[toc,page]{appendix} 
\usepackage{tabularx}
\usepackage{url}
\usepackage{enumitem}
\setlist[itemize]{leftmargin=*}

\usepackage[ruled]{algorithm2e}

\AtBeginDocument{%
  \providecommand\BibTeX{{%
    \normalfont B\kern-0.5em{\scshape i\kern-0.25em b}\kern-0.8em\TeX}}}

\copyrightyear{2024}
\acmYear{2024}
\setcopyright{acmlicensed}\acmConference[CHI '24]{Proceedings of the CHI Conference on Human Factors in Computing Systems}{May 11--16, 2024}{Honolulu, HI, USA}
\acmBooktitle{Proceedings of the CHI Conference on Human Factors in Computing Systems (CHI '24), May 11--16, 2024, Honolulu, HI, USA}
\acmDOI{10.1145/XXXXXXX.XXXXXXX}
\acmISBN{XXX-X-XXXX-XXXX-X/XX/XX}

\begin{document}

\title{``Are You Really Sure?'' Understanding the Effects of Human Self-Confidence Calibration in AI-Assisted Decision Making}

\author{Shuai Ma}
\orcid{0000-0002-7658-292X}
\affiliation{
  \institution{The Hong Kong University of Science and Technology}
  \city{Hong Kong}
  \country{China}
}
\email{shuai.ma@connect.ust.hk}

\author{Xinru Wang}
\affiliation{
  \institution{Purdue University}
  \city{West Lafayette}
  \state{Indiana}
  \country{USA}
}
\email{xinruw@purdue.edu}

\author{Ying Lei}
\affiliation{
  \institution{East China Normal University}
  \city{Shanghai}
  \country{China}
}
\email{10195102413@stu.ecnu.edu.cn}

\author{Chuhan Shi}
\affiliation{
  \institution{Southeast University}
  \city{Nanjing}
  \country{China}
}
\email{chuhanshi@seu.edu.cn}

\author{Ming Yin}
\affiliation{
  \institution{Purdue University}
  \city{West Lafayette}
  \state{Indiana}
  \country{USA}
}
\email{mingyin@purdue.edu}

\author{Xiaojuan Ma}
\affiliation{
  \institution{The Hong Kong University of Science and Technology}
  \city{Hong Kong}
  \country{China}
}
\email{mxj@cse.ust.hk}

\renewcommand{\shortauthors}{Shuai Ma, et al.}

\begin{abstract}
In AI-assisted decision-making, it is crucial but challenging for humans to achieve appropriate reliance on AI. This paper approaches this problem from a human-centered perspective, ``human self-confidence calibration''. We begin by proposing an analytical framework to highlight the importance of calibrated human self-confidence. In our first study, we explore the relationship between human self-confidence appropriateness and reliance appropriateness. Then in our second study, We propose three calibration mechanisms and compare their effects on humans' self-confidence and user experience. Subsequently, our third study investigates the effects of self-confidence calibration on AI-assisted decision-making. Results show that calibrating human self-confidence enhances human-AI team performance and encourages more rational reliance on AI (in some aspects) compared to uncalibrated baselines. Finally, we discuss our main findings and provide implications for designing future AI-assisted decision-making interfaces.

\end{abstract}




\maketitle

\input{sections/01-Introduction}
\input{sections/02-RelatedWork}
\input{sections/03-Framework}
\input{sections/04-Study1}
\input{sections/05-Study2}

\input{sections/06-Study3}

\input{sections/07-Discussion}


\bibliographystyle{ACM-Reference-Format}
\bibliography{main}


\end{document}

%% file: sections/01-Introduction.tex
\section{Introduction}
AI technology is increasingly crucial in supporting human decision-making across various domains \cite{ma2023should, buccinca2021trust, zhang2020effect, wang2021explanations, bansal2021does, zheng2023competent, zheng2024charting}. In AI-assisted decision-making, AI provides recommendations while leaving the final decision to humans \cite{zhang2020effect}. Given the inherent uncertainties of both humans and AI, one key challenge is ensuring humans' appropriate reliance on AI \cite{bansal2021does, zhang2020effect}. Showing AI confidence levels has been proposed to address this, as accurate confidence scores can indicate the likelihood of correct predictions \cite{zhao2023evaluating, zhang2020effect, ma2023should, bansal2021does, rastogi2020deciding, lai2022human}. Nonetheless, studies on AI confidence presentation show mixed outcomes, suggesting it doesn't always improve human-AI collaboration outcomes \cite{zhang2020effect, zhao2023evaluating, ma2023should}.


A key reason for the limited effectiveness of showing AI confidence is that people's reliance is not solely based on AI confidence but also their self-confidence \cite{vodrahalli2022humans, chong2022human}. For instance, overconfident individuals may dismiss correct AI recommendations, while underconfident ones may overly rely on erroneous AI advice. Existing research often overlooks the role of human self-confidence in this process, assuming that individuals possess an appropriate perception of their confidence and can make rational decisions after evaluating AI's confidence. However, extensive evidence from decision-making and cognitive science literature shows that people frequently exhibit poorly calibrated self-confidence \cite{turner2022calibrating, meyer2013physicians, weber2004confidence, moore2020perfectly, miller2015meta}.

In this work, we address this crucial issue and propose an innovative approach to improve the collaboration between humans and probabilistic AI models through \emph{human self-confidence calibration}. We first introduce an analytical framework to uncover inappropriate human reliance from a confidence-correctness matching perspective, recognizing that inappropriate self-confidence may hinder rational human reliance on AI. Then, through three consecutive studies, we aim to explore three critical research questions:

\begin{itemize}
    \item \textbf{RQ1:} How may humans' inappropriate self-confidence affect the appropriateness of their reliance on AI's suggestions?
    \item \textbf{RQ2:} How can humans' self-confidence be calibrated and how will different self-confidence calibration mechanisms affect humans' perceptions and user experience?
    \item \textbf{RQ3:} How will calibration of humans' self-confidence affect the appropriateness of their reliance on AI's suggestions and task performance?
\end{itemize}

To answer RQ1, we conduct our first study using income prediction as the task \cite{zhang2020effect, hase2020evaluating, ribeiro2018anchors, ghai2021explainable}. Following the Judge Advisor System (JAS) model\cite{sniezek1995cueing}, we performed a three-step process: individuals initially made a judgment and assessed their self-confidence, then received AI recommendations (with or without a confidence score), and ultimately made their final judgment. Our findings uncover a significant link between the appropriateness of human self-confidence and the appropriateness of human reliance on AI. Through our analytical framework, we discovered that discrepancies between human confidence and correctness significantly increase error rates. These insights underpin our core goal: \emph{calibrating individuals' self-confidence to optimize the appropriateness of their reliance on AI}.

In response to RQ2, built on decision-making and cognitive science theories related to human self-confidence calibration, we introduce three calibration mechanisms: \emph{Think the Opposite (Think)}, \emph{Thinking in Bets (Bet)}, and \emph{Calibration Status Feedback (Feedback)}. Then, in our second study, we deliberately removed AI involvement to mitigate potential confounding factors stemming from AI model suggestions. Participants independently made a set of predictions, and we compared the three proposed self-confidence calibration mechanisms against a control condition (without any calibration). Besides objective metrics, we also collected participants' perceptions and user experience. The results demonstrate that, compared to the control condition, \emph{Think} and \emph{Feedback} effectively align participants' self-confidence levels with their actual accuracy. However, \emph{Think} yields participants' higher perceived complexity and mental demand, as well as lower user preference and satisfaction. These findings imply that balancing these trade-offs will be a pivotal consideration for future research.

Finally, to answer RQ3, our third study explored the impact of self-confidence calibration on AI-assisted decision-making. We compared the results of confidence calibration with a baseline without calibration. The findings indicate that confidence calibration leads to people's more rational reliance behaviors, reduces their under-reliance (though over-reliance is not reduced), and improves task performance. In addition, based on our analytical framework, we also analyzed the proportion of different human-AI confidence-correctness matching situations and the corresponding error rates in a fine-grained manner.

In this paper, we make the following contributions:

\begin{itemize}
    \item We proposed an analytical framework that unpacks humans' (in)appropriate reliance from a novel perspective, i.e., human self-confidence appropriateness. This framework provides fresh insights for understanding reliance appropriateness.
    \item We designed an exploratory study to understand the relationship between the appropriateness of human self-confidence and the appropriateness of human reliance on AI.  
    \item We designed three mechanisms for calibrating self-confidence and evaluated their efficacy and impact on user experience while analyzing their advantages and limitations.
    \item We further investigated the influence of self-confidence calibration on human reliance on AI suggestions and decision-making performance. We furnished substantial evidence and a deep understanding of the pivotal role of self-confidence calibration in human-AI collaborative decision-making.
\end{itemize}

In summary, this paper offers a distinctive perspective on understanding and prompting the appropriateness of human reliance in AI-assisted decision-making. We aspire that our investigation will contribute to enriching the community's comprehension of the role of human self-confidence in human-AI collaboration and serve as a cornerstone for continued research of self-confidence calibration methodologies within the realm of AI-assisted decision-making.

%% file: sections/02-RelatedWork.tex
\section{Related work}

\subsection{Appropriate Reliance in AI-Assisted Decision-Making and Its Measurements}

Extensive research has examined the appropriateness of human reliance on AI systems (including broadly automated systems and robotics) \cite{ma2024beyond, wang2008selecting, lee2004trust, buccinca2021trust, wischnewski2023measuring, ma2022glancee, ma2022modeling}. In recent HCI studies, the focus has shifted from merely increasing trust in AI to facilitating appropriate trust and reliance \cite{10.1145/3579612, vasconcelos2023explanations, buccinca2020proxy, vereschak2021evaluate, zhang2020effect, schemmer2023appropriate, ma2023should, yang2023harnessing, wischnewski2023measuring}. Two widely recognized phenomena, automation bias \cite{parasuraman2010complacency} and algorithm aversion \cite{dietvorst2015algorithm}, highlight the challenge individuals face in aligning their trust and reliance with AI system capabilities.

In AI-assisted decision-making research, the focus has turned toward managing trust and reliance on a case-by-case basis \cite{vereschak2021evaluate, lai2021towards, wischnewski2023measuring}. Key concepts in this context are \emph{trust} and \emph{reliance} \cite{vereschak2021evaluate, lai2021towards, wischnewski2023measuring}. \emph{Trust} reflects subjective perceptions of AI, often assessed using self-report scales, while \emph{reliance} pertains to objective behavior in response to AI systems \cite{zhang2020effect, bansal2021does, ma2023should, zhao2023evaluating, wang2021explanations}. Intriguingly, studies have revealed inconsistencies between trust and reliance. Increased self-reported trust doesn't necessarily correlate with improved reliance behaviors \cite{rechkemmer2022confidence}. In this paper, our focus is on studying human \emph{reliance} behaviors, which often provide a more reliable indicator of appropriateness when relying on AI compared to self-reported trust.

Appropriate reliance involves accepting AI suggestions when they are correct and rejecting them when they are wrong. Existing studies employ diverse definitions and measurement methods for appropriate reliance, falling into two broad categories:

\textbf{Behavior-Based Measurement}: This approach assesses appropriate reliance by analyzing human behaviors, considering AI confidence as an indicator of trustworthiness \cite{zhang2020effect, zhao2023evaluating}. If humans rely more on AI's suggestions when the AI expresses high confidence (and rely on AI less when AI confidence is low), it's deemed calibrated and appropriate. However, the confidence of AI may not directly represent the correctness of AI. Therefore, many works that rely on this measure of the appropriateness of human reliance have found that although people's trust/reliance gets calibrated, the final task performance does not improve \cite{zhang2020effect, zhao2023evaluating}.

\textbf{Outcome-Based Measurement}: This approach directly assesses appropriate reliance based on the correctness of AI recommendations and human decisions \cite{bansal2021does, ma2023should, wang2021explanations, schemmer2023appropriate, he2023knowing}. It categorizes inappropriate reliance into over-reliance and under-reliance, quantifying the appropriateness of human reliance on AI suggestions. \textbf{Over-reliance} occurs when people align with AI predictions when the AI is incorrect, while \textbf{under-reliance} is when people reject AI predictions when the AI is correct. Some work has adopted more stringent criteria, focusing solely on whether people can make the correct final decisions when their initial predictions and AI suggestions differ \cite{he2023knowing, schemmer2023appropriate}. In this paper, we use outcome-based measurement to assess humans' reliance appropriateness.

\subsection{Enhancing Appropriate Reliance in AI-Assisted Decision-Making}

The field of AI-assisted decision-making is gaining significant attention within HCI communities \cite{vasconcelos2023explanations, buccinca2020proxy, vereschak2021evaluate, zhang2020effect, schemmer2023appropriate, ma2023should, yang2023harnessing, wischnewski2023measuring, zhu2022bias, shi2023retrolens, ma2019smarteye}. In this human-AI collaborative setting, AI acts as an advisor, offering suggestions often accompanied by uncertainty. A paramount challenge in these scenarios is ensuring that humans rely on AI advice in an appropriate manner \cite{zhang2020effect, lai2021towards}. To tackle this issue, existing studies have ventured into three primary strategies.

One approach centers on enhancing individuals' comprehension of AI prediction uncertainty \cite{zhang2020effect, wischnewski2023measuring}. AI prediction uncertainty is typically measured via the AI model's calibrated confidence level that can reflect prediction correctness probabilities \cite{rechkemmer2022confidence, guo2017calibration}. For example, a confidence score of 0.6 signifies a 60\% chance of a correct prediction. Some studies directly display calibrated confidence scores to users \cite{zhang2020effect, bansal2021does}, which explored the impact of showing AI confidence on trust calibration and task performance. Others integrate AI confidence into interface design. For instance, Rastogi et al. \cite{rastogi2020deciding} adjusted decision-making timeframes based on AI confidence levels to promote humans' more analytical thinking when AI's confidence is low. Besides, various confidence representation methods, including violin plots or question marks, have also been explored \cite{zhao2023evaluating}. However, these approaches haven't consistently resulted in improved reliance appropriateness or task performance \cite{zhang2020effect, ma2023should, zhao2023evaluating} - ``only displaying AI confidence can be insufficient''.

The second approach focuses on enhancing individuals' understanding of AI error patterns, aiding in the development of humans' accurate mental models for AI capabilities \cite{he2023interaction, bansal2019beyond, 10.1145/3579612}. For instance, Bansal et al. \cite{bansal2019beyond} introduced the concept of ``mental models of AI error boundaries'', highlighting factors shaping these models. This enables individuals to discern when to accept or reject AI recommendations. Cabrera et al. \cite{10.1145/3579612} proposed to display ``behavior descriptions'' of AI models to end-users, providing insights into AI performance on specific instances. This approach enhances human-AI collaboration by helping users recognize AI failures and fostering more reliance on AI when it demonstrates higher accuracy.

The third approach aims to elucidate the rationale behind AI predictions through AI explanations \cite{he2023knowing, vasconcelos2023explanations, bansal2021does, wang2019designing, poursabzi2021manipulating, lai2020chicago, lai2019human}. These explanations take various forms, such as feature importance, feature contribution, similar examples, counterfactual examples, and natural language-based explanations \cite{lai2021towards, liao2021human}. However, recent research has unveiled a potential drawback of providing AI explanations: the risk of increased over-reliance on AI systems when AI provides incorrect suggestions \cite{bansal2021does, poursabzi2021manipulating, wang2021explanations, zhang2020effect}. This phenomenon is attributed to a lack of cognitive engagement with AI explanations, as individuals may opt for quick heuristic judgments, associating explainability with trustworthiness when they lack the motivation or ability for in-depth analysis \cite{bertrand2022cognitive, buccinca2020proxy}.

Our approach, distinct from prior methods, focuses on enhancing the appropriateness of humans' reliance by calibrating their self-confidence. One similar work to ours is He et al.'s study \cite{he2023knowing}, which addresses individuals' overestimation of their abilities (known as the Dunning-Kruger effect) by calibrating self-assessment through a tutorial. However, their approach primarily targets task-level self-assessment, whereas rational reliance requires case-by-case judgments on whether to adopt AI recommendations \cite{zhang2020effect}. Moreover, they didn't explore the setting where AI shows confidence, whereas our work studies the effects of calibrating human self-confidence when AI's confidence is also presented. Another related work by Ma et al. \cite{ma2023should} models the correctness likelihood (CL) of humans and AI, comparing them within each task case to adaptively adjust the decision-making interface. However, their primary emphasis was on enhancing AI's understanding of humans, rather than individuals gaining self-calibration. Additionally, they found some users doubted the system-estimated human CL which hindered the effectiveness of their approach. Besides, they nudged user choices through the interface design, potentially compromising user autonomy and raising ethical concerns. In contrast, our work focuses on calibrating individual self-confidence, not only ensuring user autonomy but also avoiding ethical issues around AI nudges.

\subsection{Human Self-confidence in Decision Making and the Calibration}

Confidence, grounded in subjective perceptions, shapes our belief in the validity of our thoughts and abilities \cite{luttrell2013metacognitive, grimaldi2015there}. It plays a pivotal role in decision-making and receptiveness to advice \cite{bonaccio2006advice}, even affecting our willingness to heed AI recommendations \cite{vodrahalli2022humans, chong2022human, lu2021human}. Humans' self-confidence often correlates with credibility in various contexts, from children's perceptions of adults \cite{tenney2011accuracy} to juror evaluations of expert witnesses \cite{cramer2011confidence}. However, self-confidence can sometimes stray from reality, leading to overconfidence or underconfidence, affecting experts and laypersons alike \cite{turner2022calibrating, meyer2013physicians, weber2004confidence, moore2020perfectly, miller2015meta}. Overconfidence, characterized by inflated self-estimation, can result in risky choices \cite{moore2008trouble}. Conversely, underconfidence, marked by self-underestimation, can lead to missed opportunities \cite{kruger1999unskilled, dunning2003people}. Extensive empirical studies in decision-making have observed the misalignment between human self-confidence and actual accuracy, evident in clinical diagnosis and financial decisions \cite{miller2015meta, grevzo2021overconfidence}. For example, Miller et al. \cite{miller2015meta} found no consistent correlation between clinicians' confidence and decision accuracy, while Grežo et al. \cite{grevzo2021overconfidence} revealed overconfidence's significant impact on financial decisions. These findings highlight the importance of accurate self-assessment.


The literature on self-confidence calibration delves into cognitive processes and mechanisms. This research uncovers cognitive biases and heuristics contributing to mis-calibration, such as the impact of overconfidence \cite{moore2008trouble} and the Dunning-Kruger effect \cite{kruger1999unskilled}. It also explores metacognitive processes, offering manipulations to enhance calibration \cite{koriat2006illusions, pleskac2010two}.
To foster calibrated self-confidence, cognitive approaches have emerged. Pulford et al. \cite{pulford1997overconfidence} examine external feedback and response time's impact on calibration. Moore, in his book ``Perfectly Confident'' \cite{moore2020perfectly}, navigates human confidence complexities, highlighting factors influencing accurate judgments and offering practical strategies, including seeking feedback, diverse perspectives, and a growth mindset. Duke \cite{duke2019thinking} advocates embracing uncertainty and viewing decisions as bets, offering a strategy to assess risks, uncertainties, and potential outcomes.

Our paper adapted human self-confidence calibration to AI-assisted decision-making scenarios, examining how it influences humans' reliance on AI suggestions amid uncertainty.

%% file: sections/03-Framework.tex
\section{Unpacking Inappropriate Reliance from a Human Self-Confidence Perspective}

\subsection{Appropriateness of Human Self-Confidence}
The appropriateness of human self-confidence depends on how well it aligns with actual competence or performance \cite{moore2008trouble, moore2017confidence}. Overconfidence happens when confidence exceeds abilities, while underconfidence occurs when confidence falls short. In decision-making research, evaluating self-confidence appropriateness involves gathering humans' predictions and corresponding self-reported confidence levels \cite{turner2022calibrating, meyer2013physicians, weber2004confidence, moore2020perfectly}. Next, we introduce the measurements of confidence appropriateness at both task and instance levels.

\subsubsection{Existing Measurements of Confidence Appropriateness at A Task Level}

\begin{figure}[t]
	\centering 
	\includegraphics[width=\linewidth]{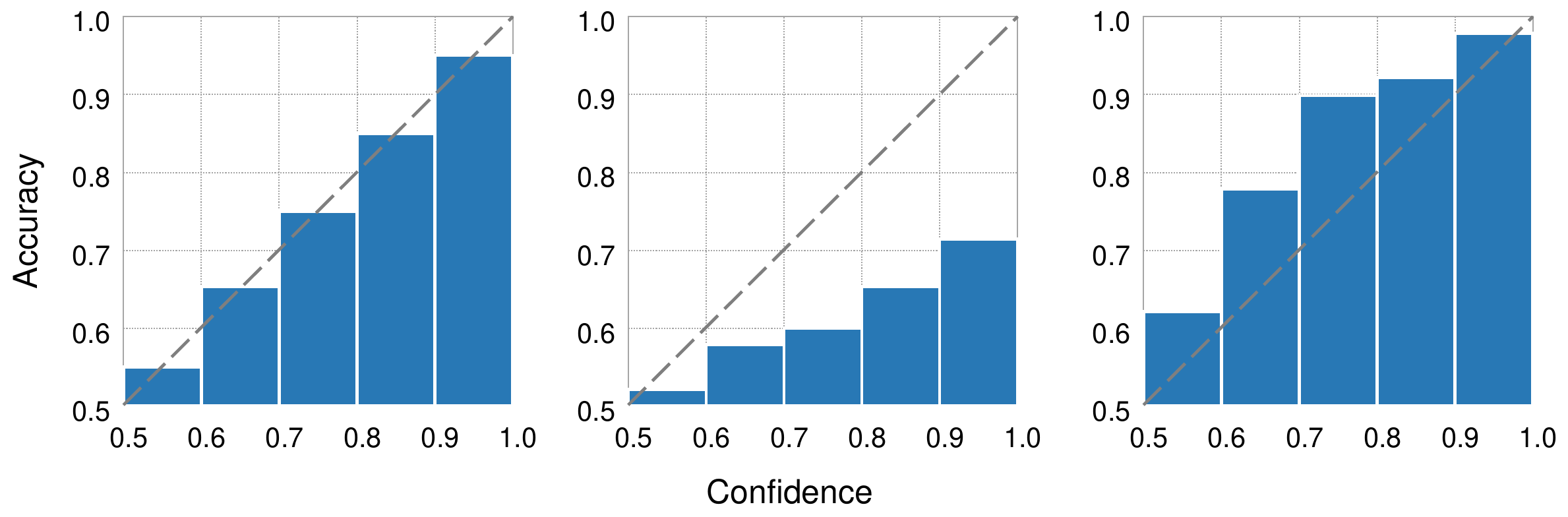}
	\caption{Reliability diagrams for a binary classification task \cite{guo2017calibration}, illustrating calibrated confidence (left, the actual accuracy aligns with the stated confidence), over-confidence (middle, the actual accuracy falls below the stated confidence), and under-confidence (right, the actual accuracy is above the stated confidence).}
        \Description{This figure shows three bar charts, each with an x-axis representing people's stated confidence and a vertical axis representing true accuracy. In each chart, there are 5 bars and a 45-degree dashed line representing y=x. In the leftmost chart, the 5 bars are exactly on the 45-degree dashed line representing y=x. In the middle chart, the 5 bars are all lower than the 45-degree dashed line representing y=x. In the right chart, the 5 bars are all higher than the 45-degree dashed line representing y=x.}
	\label{fig:reliability}
\end{figure}

Many measurements have been proposed to evaluate the appropriateness of confidence at a task level, such as Over/Under Confidence Index \cite{weber2004confidence, meyer2013physicians}, Brier score \cite{rufibach2010use}, Pearson correlation coefficient \cite{depaulo1997accuracy, miller2015meta, rechkemmer2022confidence}, etc. One of the most widely used measurements is Reliability diagrams \cite{degroot1983comparison, niculescu2005predicting, hartmann2002confidence} (Figure \ref{fig:reliability}), assessing the alignment between stated confidence and actual accuracy.

For a task, reliability diagrams first partition all $N$ predictions into $M$ bins based on the corresponding confidence values, then calculate the accuracy $acc(B_m)$, and average confidence $conf(B_m)$, for each bin $B_m$. Finally, the diagrams can be drawn by setting confidence as the horizontal axis and actual accuracy as the vertical axis. With the diagrams, a metric called Expected Calibration Error (ECE) \cite{weber2004confidence, meyer2013physicians} is used to quantify the difference between expected accuracy and self-reported confidence over the partitioned bins.

\begin{small}
\begin{equation}
    \textbf{ECE} = \sum_{m=1}^{M}\frac{|B_{m}|}{N}|acc(B_{m} ) - conf(B_{m})|,
\label{equation1}
\end{equation}
\end{small}

ECE first computes the absolute difference between the accuracy $acc(B_m)$ and average confidence $conf(B_m)$ within each bin $B_m$, then calculates the average of all bins, weighted by the number of predictions of each bin $|B_m|$ over the total prediction number $N$. In our user studies, we will use ECE to measure the overall appropriateness of human self-confidence. Since appropriate reliance requires humans to distinguish whether to rely on an AI's suggestion on a \textbf{case-by-case} basis \cite{zhang2020effect, schemmer2023appropriate}, to understand the effects of self-confidence appropriateness on reliance appropriateness, we should also measure it at an instance level. Thus, next, we propose an instance-level measurement of self-confidence appropriateness.

\renewcommand{\arraystretch}{1.5}
\begin{table}[htp]  

\centering  
\fontsize{8}{8}\selectfont 

\caption{Instance-Level Confidence-Correctness Matching. There are four types of confidence \& correctness situations related to a specific prediction in decision-making tasks. Whether a C-C is matched does not depend on whether the prediction is correct, but on whether the correctness is aligned with confidence.}
\begin{tabular}{ccc}
\hline
\textbf{Confidence} & \textbf{Correctness} & \textbf{C-C Matching}\\
\hline
 High & Correct & C-C Matched \\
High & Incorrect & Over-confident (C-C Mismatched) \\
Low & Correct & Under-confident (C-C Mismatched) \\
Low & Incorrect & C-C Matched \\
\hline
\end{tabular} 
\label{alignment}
\end{table}

\subsubsection{Measuring Confidence Appropriateness at An Instance Level}
\label{sec3.1.2}
Based on the confidence level and the correctness of a specific prediction, we propose a measurement called \textbf{Confidence-Correctness Matching} (\textbf{C-C Matching} in short, shown in Table \ref{alignment}). To simplify the problem, in this paper, we consider confidence at a binary level: low or high\footnote{Note that the threshold of confidence levels (high or low) depends on some factors such as task characteristics \cite{rastogi2020deciding, rechkemmer2022confidence, lu2021human, bansal2021does}. For instance, for binary classification tasks, AI confidence values fall within the range of 0.5 to 1.0, while for multi-classification tasks, this range may extend from 0 to 1.0. Some previous studies have employed thresholds (e.g., mean or median) to define what constitutes "high" confidence \cite{rastogi2020deciding, rechkemmer2022confidence, lu2021human, bansal2021does}}. For a classification task, any prediction can be categorized into four types based on its confidence (low or high) and correctness (correct or incorrect). We define [High confidence \& Incorrect prediction] as \emph{Over-confident} and [Low confidence \& Correct prediction] as \emph{Under-confident} in a specific prediction. And we classify these two as \textbf{Confidence-Correctness Mismatched} (C-C Mismatched in short). On the contrary, we define [High confidence \& Correct prediction] and [Low confidence \& Incorrect prediction] as \textbf{Confidence-Correctness Matched} (C-C Matched in short). Based on C-C Matching, we propose an analytical framework to analyze the appropriateness of humans' reliance on AI.

\begin{figure*}[t]
	\centering 
	\includegraphics[width=0.9\linewidth]{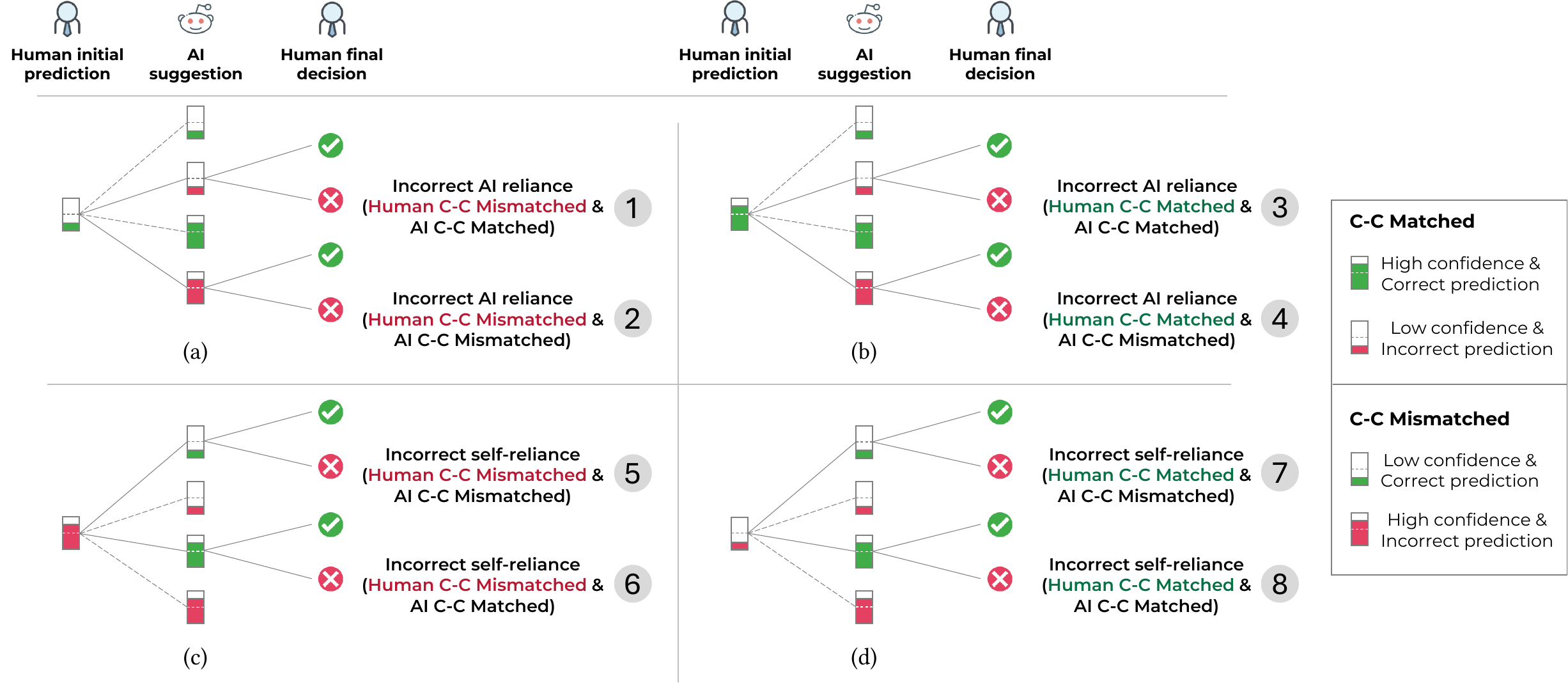}
	\caption{A space of different combinations of 1) initial human prediction correctness and confidence, 2) AI suggestion correctness and its confidence, and 3) human final decision correctness, at a task instance level. To save space, we only highlight situations where a human's initial prediction differs from the AI's suggestion and the human's final decision is incorrect. Comparing (a) and (b), (a) may induce more incorrect AI reliance due to \emph{Human C-C Mismatched} (Low\&Correct). Similarly, (c) may lead to more incorrect self-reliance due to \emph{Human C-C Mismatched} (High\&Incorrect).}
        \Description{This figure has 2 * 2 subgraphs. Each subgraph shows different situations of human and AI's confidence-correctness matching. High confidence \& correct prediction is represented by a green measuring cup with a full fill. High confidence \& incorrect prediction is represented by a red measuring cup with a full fill. Low confidence \& correct prediction is represented by a green measuring cup with less filling. Low confidence \& incorrect prediction is represented by a red measuring cup with less filling. The subgraph in the upper left corner depicts a green measuring cup with less filling for humans, two types of red measuring cups for AI, and the final prediction of errors in both cases. The subgraph in the upper right corner depicts a green measuring cup with full filling for humans, two types of red measuring cups for AI, and the final prediction of errors in both cases. The subgraph in the lower left corner depicts a red measuring cup with full filling for humans, The two green measuring cups of AI, as well as the final predictions of errors in both cases. The subgraph in the lower right corner depicts the red measuring cup with less filling for humans, the two green measuring cups of AI, and the final predictions of errors in both cases.}
	\label{fig:analysisframework}
\end{figure*}

\subsection{An Analytical Framework Integrating Human and AI Confidence Appropriateness}

Existing studies on improving reliance appropriateness in human-AI decision-making often focus on the AI confidence perspective (e.g., providing different forms of AI confidence) \cite{zhang2020effect, zhao2023evaluating, rastogi2020deciding}. However, they overlook the significance of assessing the appropriateness of human self-confidence \cite{wang2021explanations, bansal2021does}. Within the context of human-AI collaborative decision-making, the \textbf{interplay} between individuals' confidence in their own judgments and the confidence expressed by AI systems plays a pivotal role in shaping human reliance on AI recommendations \cite{vodrahalli2022humans, chong2022human}. Therefore, to comprehensively investigate and analyze the intricate relationship among these factors, we propose an integrated analytical framework. This framework takes the \textbf{Confidence-Correctness Matching} (see Sec \ref{sec3.1.2}) of both humans and AI into consideration to analyze the specific causes of inappropriate reliance.

We adopt the Judge-Advisor System (JAS) in decision-making to build our analytical framework (Figure \ref{fig:analysisframework}). In JAS, there will be three types of predictions: (1) human initial prediction, (2) AI suggestion, and (3) human final decision. Figure \ref{fig:analysisframework} illustrates all combinations of human and AI C-C Matching situations. For clarity, here, we focus on cases where humans initially disagree with AI (because humans tend to keep their initial predictions when they are the same as the AI’s recommendations, feeling confirmed in their choices \cite{bansal2021does, schemmer2023appropriate}). Next, we analyze the potential causes of inappropriate reliance based on humans and AI's C-C Matching.

\textbf{From Human C-C Matching Perspective.} Within Figure \ref{fig:analysisframework}, we distinguish two types of incorrect reliance: incorrect AI reliance (Figure \ref{fig:analysisframework} (a) and (b)) and incorrect self-reliance (Figure \ref{fig:analysisframework} (c) and (d), which is also another perspective of incorrect AI reliance). However, we speculate that the causes of incorrect AI reliance in Figure \ref{fig:analysisframework} (a) may be different from Figure \ref{fig:analysisframework} (b). Similarly, the causes of incorrect self-reliance in Figure \ref{fig:analysisframework} (c) can be different from Figure \ref{fig:analysisframework} (d). Specifically, for humans' incorrect AI reliance in Figure \ref{fig:analysisframework} (a) and (b): although in both cases, the human first makes a correct initial prediction and then sees a wrong AI suggestion, the Human C-C Matching is different. In (a), people’s confidence in their initial judgment is low (low \& correct, C-C Mismatched), but in (b), people are very confident in their initial judgment (high \& correct, C-C Matched). We conjecture that humans in (a) are more prone to adopt the AI's erroneous suggestions due to their low self-confidence. Similar analysis can be used for humans' incorrect self-reliance in Figure \ref{fig:analysisframework} (c) and (d). We speculate that humans in case (c) are more prone to ignore the AI's correct suggestions due to their mistakenly high confidence levels.

\textbf{From AI C-C Matching Perspective.} If AI is C-C Mismatched, it can pose challenges for a human to appropriately rely on AI's suggestions. For instance, Figure \ref{fig:analysisframework} (1)-(4) are all situations where humans with a correct initial prediction encounter an incorrect AI suggestion. However, the AI C-C Matching varies. In situations (2) and (4), the AI provides an incorrect suggestion but with high confidence (AI C-C Mismatched). Compared to situations (1) and (3), in situations (2) and (4), humans could easily be misled by the AI's high confidence, resulting in incorrect AI reliance. Similarly, Figure \ref{fig:analysisframework} (5)-(8) are all situations where humans with an incorrect initial prediction encounter a correct AI suggestion. However, in situations (5) and (7), the AI provides a correct suggestion but with low confidence (AI C-C Mismatched). Compared to situations (6) and (8), humans in situations (5) and (7) could be more likely to ignore AI's correct suggestions due to AI's low confidence, leading to incorrect self-reliance.

Overall, we argue that Human and AI Confidence-Correctness Matching jointly influences the appropriateness of human reliance. If both humans and AI are C-C Mismatched (Figure \ref{fig:analysisframework} (2) and (5)), it can be extremely challenging for humans to achieve appropriate reliance. Conversely, if both humans and AI are C-C Matched (Figure \ref{fig:analysisframework} (3) and (8)), humans would be more likely to have correct reliance. While the AI community has explored calibrating AI confidence to enhance AI C-C Matching \cite{guo2017calibration}, scant focus in the HCI community has been given to human confidence calibration and little is known about its impact on reliance appropriateness. To fill this research gap, this paper introduces a self-confidence calibration method designed to improve Human C-C Matching. Through this calibration approach, we aim to reduce the occurrence of Human C-C Mismatch, ultimately mitigating incorrect reliance stemming from such discrepancies.

\subsubsection{How can we use the proposed analytical framework?}

\textbf{Helping with Posthoc Analysis of Inappropriate Reliance.} One crucial application of this framework is its use in dissecting the causes behind people's inappropriate reliance, from a Confidence-Correctness Matching perspective. Specifically, we can categorize users' decision-making data into different human-AI C-C Matching situations. By checking the occurrence ratio of each situation, we can know whether humans or AI have confidence-related problems.

\textbf{Informing AI System Design.} The detailed understanding of the causes of users' inappropriate reliance can further enable designers to make targeted enhancements to AI system design. For instance, if inappropriate reliance predominantly stems from frequent AI C-C Mismatch, designers can involve mechanisms to refine the calibration of the AI model's confidence. Conversely, if the root cause lies in recurring Human C-C Mismatch, designers can add interventions to calibrate users' self-confidence to improve their rationality in the decision-making process.

%% file: sections/04-Study1.tex
\section{Study 1 - Understanding the Relationship between Human Self-Confidence Appropriateness and Reliance Appropriateness}

Our first study aims to understand the relationship between the appropriateness of human self-confidence and the appropriateness of human reliance. In this study, we did not perform any intervention on the participants' self-confidence to capture their most natural behaviors when making decisions with AI's assistance.

\subsection{Research Questions}
Focusing on our main research question \textbf{RQ1: How may humans' inappropriate self-confidence affect their reliance appropriateness on AI's suggestions?}, we specifically ask the following sub-questions.

As mentioned in our analytical framework (Sec 3.2), inappropriate human self-confidence (C-C Mismatched) might affect reliance appropriateness. Therefore, we first ask,
\begin{itemize}
    \item[-] \textbf{RQ 1.1}: How will different situations of human \emph{C-C Matching} affect humans' performance?
    \item[-] \textbf{RQ 1.2}: How will the appropriateness of human self-confidence correlate with the appropriateness of human reliance?
\end{itemize}

In addition, before calibrating humans' self-confidence, we want to first explore whether there will be any difference when AI's confidence is shown or not.
\begin{itemize}
    \item[-] \textbf{RQ 1.3}: How will the presence of AI confidence affect the appropriateness of human self-confidence?
    \item[-] \textbf{RQ 1.4}: How will the presence of AI confidence affect the appropriateness of human reliance and task performance?
\end{itemize}

\subsection{Task and AI Model}

\subsubsection{Task}
We selected \emph{income prediction} as our testbed, which is widely used in existing AI-assisted decision-making studies \cite{ma2023should, zhang2020effect, hase2020evaluating, ribeiro2018anchors, ghai2021explainable}. Participants were tasked with predicting whether an individual's annual income exceeded \$50K based on her/his profile. Data for this task came from the Adult Income dataset \cite{datasetucl} in the UCI Machine Learning Repository, comprising 48,842 instances with 14 attributes. The ground truth was binary (greater/less than 50K). We chose income prediction as our task for three reasons. First, it does not require specific domain knowledge or training which is suitable for non-expert participants \cite{ghai2021explainable}. Second, the task is relatively low-risk so factors such as personal risk tolerance and responsibility concerns have less influence on people’s reliance on AI, allowing us to focus on studying the effects of human-AI confidence. Third, prior research suggests that in the income prediction task, lay people's confidence can sometimes be poorly calibrated \cite{ma2023should}. This makes it an ideal testbed for us to investigate the effects of confidence calibration. We followed the approach of \cite{zhang2020effect, ghai2021explainable}, selecting eight important attributes to present to participants, including \emph{Age}, \emph{Year of education}, \emph{Work class}, \emph{Occupation}, \emph{Marital status}, \emph{Gender}, \emph{Race}, and \emph{Work hours per week}.

\subsubsection{AI Model}
We utilized a logistic regression (LR) model with default \emph{sklearn}\footnote{https://scikit-learn.org/} settings for our income prediction task, in line with \cite{ghai2021explainable}. The LR model optimizes the Log loss and provides a well-calibrated confidence score \cite{guo2017calibration, platt1999probabilistic}, which can avoid confounding factors caused by AI's miscalibrated confidence and help us focus on human confidence calibration. After data pre-processing, we trained our model using a 70\% random split of the dataset, while participants received prediction trials from the remaining 30\%.

\subsubsection{Task Sample Selection}
To ensure a reasonable study duration, we selected 20 task instances for the main task and incorporated additional instances for the tutorial. Our selection criteria prioritized maintaining both fidelity in data distribution \cite{wang2021explanations} and well-calibrated AI confidence scores \cite{wang2021explanations, zhang2020effect}. Within the 20 main task instances, half featured AI confidence scores below 0.75, indicating low AI confidence cases (with an average score of 0.6). Among these, six were accurately predicted by AI, resulting in a $60\%$ accuracy. The remaining half showcased confidence scores above 0.75, signifying high AI confidence cases (with an average score of 0.9), and nine of these were correctly predicted by AI, yielding a $90\%$ accuracy. We set different AI accuracies for the low-confidence samples and high-confidence samples separately because we need to ensure that not only is the overall AI model calibrated, but the confidence of the AI model on our selected task samples is also well-calibrated.

\subsection{Conditions}
To understand the relationship between the appropriateness of human self-confidence and the appropriateness of their reliance on AI, we use a natural AI-assisted decision-making process. Since we also want to explore the effects of the presence of AI confidence, we have two conditions:

\begin{itemize}
    \item \textbf{With AI Confidence}: Participants receive AI's predictions along with AI's confidence scores.
    \item \textbf{Without AI Confidence}: Participants receive only AI's predictions.
\end{itemize}

\begin{figure*}[htbp]
	\centering 
	\includegraphics[width=0.8\linewidth]{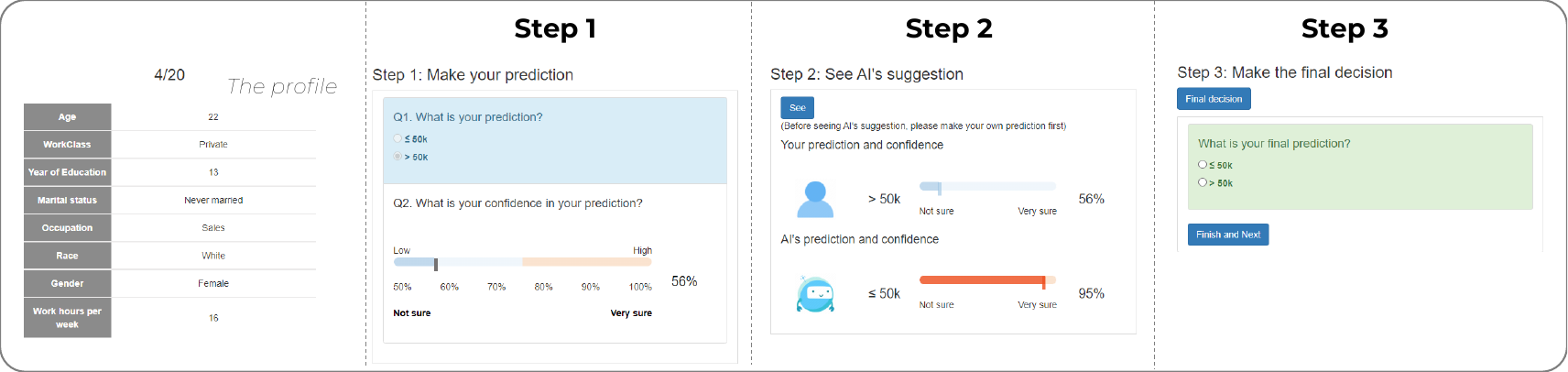}
	\caption{The interface and procedure for making a prediction on a task instance.}
	\label{fig:threestep}
         \Description{This figure shows the interface of the user decision-making process in Study 1. It is divided into four subgraphs. The leftmost subgraph shows the profile to be predicted. The second subgraph shows step 1, where people need to check the prediction radio button and drag the slider to indicate confidence. The third subgraph shows step 2, where people see their own predictions and the self-confidence represented by a slidebar, as well as the predictions of AI and the confidence of AI represented by another slidebar. The figure on the far right shows step 3, where people need to check the final prediction radio button.}
\end{figure*}

\subsection{Procedure}
After obtaining participant consent, we conducted a tutorial to familiarize them with the task. We explained each attribute in the profile table, provided income distribution graphs, and tested their understanding with qualification questions. Participants proceeded to two training examples with ground truth before the main task with AI assistance. During the main task (20 cases), participants went through three steps in each case (Figure \ref{fig:threestep}). Step 1: Participants made predictions and indicated their confidence on a slider (50\% to 100\%). And we told participants ``\emph{In a binary-choice task if you believe your confidence was lower than 50\%, you might want to flip your prediction}''. Step 2: They then received AI suggestions (with or without AI confidence). Step 3: They made final decisions. Attention-check questions were included during the main task to filter out inattentive participants.

\subsection{Participants}

Before recruiting, we performed a power analysis to determine the necessary sample size for our two-group study using G*Power \cite{faul2009statistical}. Based on a pilot study, we set the default effect size $f$ = 0.6 (indicating a moderate effect), a significance threshold $\alpha$ = 0.05, and a statistical power ($1 - \beta$) = 0.8, resulting in a sample size of 90. Following IRB approval, participants were recruited from Prolific\footnote{www.prolific.co\label{prolific}}, meeting criteria such as U.S. residency for income prediction tasks, over 99\% approval rate, English fluency, at least 1000 prior approvals, and desktop computer use. Our study, employing a between-subjects design without repeat participation, yielded 94 valid responses (\emph{With AI Confidence}: 50, \emph{Without AI Confidence}: 44) after excluding inattentive participants. Demographics included 49 males, 45 females, and varied ages and AI expertise levels. Incentives included a \$1 bonus for over 90\% accuracy. The study lasted about 15 minutes, paying an average of \$10.5 per hour.

\subsection{Evaluation Measures and Analysis}
\subsubsection{Measurements}
This study measures the appropriateness of participants' self-confidence, the appropriateness of their reliance, and their task accuracy.

\textbf{Appropriateness of human self-confidence.} We measure the Expected Calibration Error (ECE) of participants' prediction, which has been described in Eq. \ref{equation1}.

\textbf{Appropriateness of human reliance.} We employ two metrics: (1) Over-Reliance and (2) Under-Reliance.

\begin{footnotesize}
$$\textnormal{\textbf{Over-Reliance}} = \frac{\textnormal{Number of incorrect human final decisions with incorrect AI advice}}{\textnormal{Total number of incorrect AI advice}},$$
$$\textnormal{\textbf{Under-Reliance}} = \frac{\textnormal{Number of incorrect human final decisions with correct AI advice}}{\textnormal{Total number of correct AI advice}},$$

\end{footnotesize}

Based on our analytical framework (see Figure \ref{fig:analysisframework}), for the \emph{With AI Confidence} condition, we also categorize participants' predictions that initially disagree with AI into different human-AI C-C Matching situations ((1) Human C-C Mismatched \& AI C-C Matched, (2) Human C-C Matched \& AI C-C Mismatched, (3) Human C-C Mismatched \& AI C-C Mismatched, and (4) Human C-C Matched \& AI C-C Matched). We then calculated the error rate of participants' final predictions in different human-AI C-C Matching situations. 

\begin{footnotesize}
$$\textnormal{\textbf{Error Rate by C-C Matching}} = \frac{\text{number of incorrect predictions in a specific situation}}{\text{number of all predictions in a specific situation}},$$
\end{footnotesize}

\subsubsection{Analysis Method}
Since the data did not pass the normality test, we compared two unpaired groups (\emph{With AI Confidence} vs. \emph{Without AI Confidence}) via Mann-Whitney U tests and compared two paired groups (\emph{Human C-C Mismatched} vs. \emph{Human C-C Matched}) via Wilcoxon Signed-Rank tests. And we use the Spearman correlation test to analyze the correlation between the appropriateness of human self-confidence and the appropriateness of human reliance.

\subsection{Results}

\subsubsection{Effects of Different Human Confidence-Correctness Matching (RQ 1.1)}

\begin{figure*}[htbp]
	\centering 
	\includegraphics[width=0.75\linewidth]{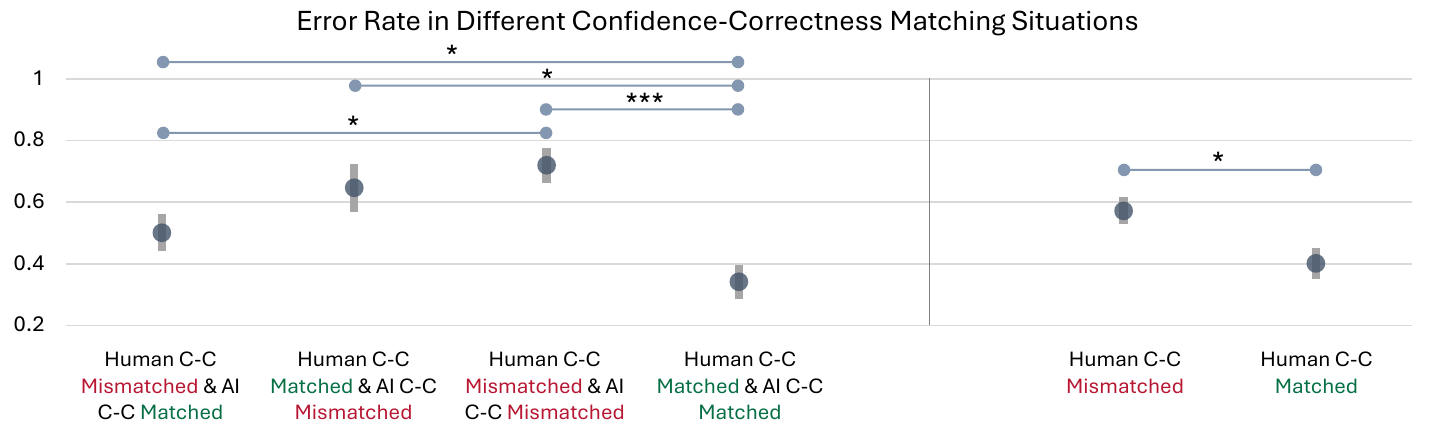}
	\caption{An analysis of error rate in different human and AI Confidence-Correctness Matching situations. The left shows the four categories considering both human and AI C-C Matching. The right shows the two categories only considering human C-C Matching no matter whether AI is C-C Matched or not. Error bars indicate standard errors. (*: $p$ < 0.05; **: $p$ < 0.01; ***: $p$ < 0.001)}
          \Description{This figure shows 6 scatter points with error bars from left to right. The leftmost point at around 0.5 represents Human C-C Mismatched\&AI C-C Matched. The second point is around 0.65, representing Human C-C Matched\&AI C-C Mismatched. The third point is around 0.75, representing Human C-C Mismatched\&AI C-C Mismatched. The fourth point is around 0.3, representing Human C-C Matched\&AI C-C Mismatched. The fifth point is around 0.6, representing Human C-C Mismatched. The sixth point is around 0.4, representing Human C-C Matched.}
	\label{fig:study1_confirmation}
\end{figure*}

Based on the proposed framework, we calculated the error rates in different human-AI C-C matching situations. Additionally, only considering human C-C matching (regardless of whether AI's confidence matched its correctness), we divided participants' task instances into two categories: (1) \emph{Human C-C Mismatched} and (2) \emph{Human C-C Matched}. Figure \ref{fig:study1_confirmation} shows the results.


Considering human and AI C-C matching together (Figure \ref{fig:study1_confirmation} left), \emph{Human C-C Mismatched \& AI C-C Matched} has a higher error rate than \emph{Human C-C Matched \& AI C-C Matched}. This indicates that when the AI's confidence matches its correctness, a mismatch in humans' confidence and correctness will lead to increased incorrect reliance. Additionally, results show that \emph{Human C-C Mismatched \& AI C-C Mismatched} yields a higher error rate than \emph{Human C-C Matched \& AI C-C Matched}. It reveals that if humans and AI both mistakenly quantify their confidence, it is \textbf{extremely hard} for humans to make a correct final decision. Only focusing on the human C-C matching (Figure \ref{fig:study1_confirmation} right), we can see that \emph{Human C-C Mismatched} showcases a higher error rate than \emph{Human C-C Matched}. This indicates that no matter whether AI's confidence matches its correctness, if the human's self-confidence is inappropriate (C-C Mismatched), the human will have more incorrect reliance.

Overall, these results validate our analytical framework's assertion that mismatches between an individual's self-confidence and actual correctness lead to increased incorrect reliance. Hence, this further supports our initial motivation - \emph{If we can calibrate people's self-confidence, we may be able to further reduce the occurrence of incorrect reliance}.

\subsubsection{Correlation between Human Self-Confidence Appropriateness and Human Reliance Appropriateness (RQ 1.2)}
We integrated \emph{With AI Confidence} and \emph{Without AI Confidence} data and conducted Spearman correlation analysis between \emph{ECE} and \emph{Under-Reliance}, and between \emph{ECE} and \emph{Over-reliance}. The results indicate that \emph{ECE} positively correlates with \emph{Under-reliance} ($\rho$: 0.404, $p$<0.001) and \emph{Over-reliance} ($\rho$: 0.343, $p$<0.01).
These findings highlight the potential for calibrating human self-confidence (lowering \emph{ECE}) to improve the appropriateness of human reliance on AI.


\subsubsection{The Effects of Showing AI Confidence (RQ 1.3, RQ 1.4)}
Our results show that there is no significant difference between with or without AI confidence in \emph{ECE}. 
Moreover, there is no significant difference in terms of \emph{accuracy}. Participants in \emph{With AI Confidence} have a higher \emph{Under-Reliance} and a lower \emph{Over-Reliance} than in \emph{Without AI Confidence}. This might be because showing AI confidence makes participants recognize the uncertainty behind AI's suggestions, leading to reduced reliance. In summary, showing AI confidence cannot improve the appropriateness of humans' self-confidence, task performance, and reliance appropriateness (at least in this paper's setting). In our Study 3, we consistently displayed AI confidence in the AI-assisted decision-making process, aiming to understand how calibration of human self-confidence will affect the decision-making outcomes when both human and AI's confidence are presented (so that humans can compare them).

\subsubsection{Summary}
In general, the results of Study 1 show that (RQ 1.1) Human C-C Mismatch will lead to more human incorrect reliance (higher error rate), so reducing the occurrence of Human C-C Mismatch has the potential to reduce humans' incorrect reliance. In addition, 
we also observed that (RQ 1.2) ECE has a strong correlation with Over-Reliance and Under-Reliance, which means that the reduction of ECE has the potential to reduce Over-Reliance and Under-Reliance. Furthermore, 
our findings indicate that (RQ 1.3 and 1.4) displaying AI confidence did not result in a reduction of ECE, nor did it directly enhance task performance or the appropriateness of human reliance on the AI system. Given the potential benefits of enhancing the appropriateness of human self-confidence, in the next study, we proceed to develop mechanisms for calibrating human self-confidence.

%% file: sections/05-Study2.tex
\section{Study 2 - Comparing the Effects of Different Self-Confidence Calibration Mechanisms}
In our second study, we aim to explore the mechanisms to calibrate human self-confidence and assess their influence on humans.

\subsection{Design of Self-Confidence Calibration}
Based on the theory and practice in cognitive science and decision-making, we propose three self-confidence calibration designs.

\begin{figure*}[htbp]
	\centering 
	\includegraphics[width=0.8\linewidth]{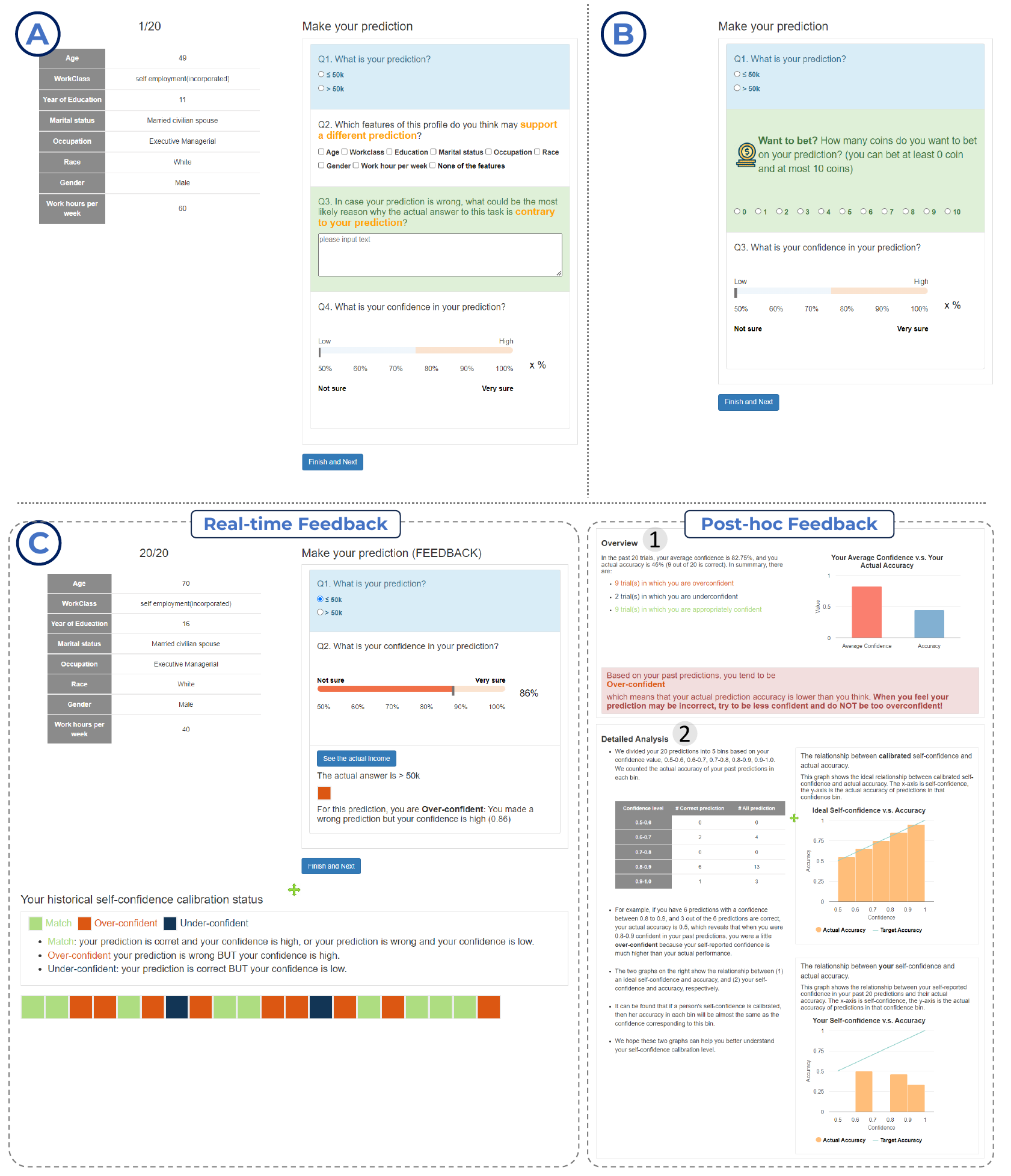}
	\caption{Interfaces of different self-confidence calibration conditions. (A) Think the Opposite. (B) Thinking in Bets. (C) Calibration Status Feedback contains two views, (1) real-time feedback during the decision-making process and (2) post-hoc feedback after a batch of decision tasks.}
           \Description{This figure shows the interface diagrams of the three calibration mechanisms used in Study 2. The top left is the Think interface, where two questions need to be answered before people indicate confidence. The top right is the Bet interface, where people need to choose how many coins to bet before indicating confidence. The bottom is the Feedback interface, whereas the bottom left is the Real-time Feedback interface, which displays a profile area, a user decision area, and a visualization area. The visualization area contains a series of square grids, divided into three colors. Green represents Match, blue represents Underconfidence, and red represents Overconfidence. In the current interface, all three colors appear, but there is more red. The bottom right is the interface of the Post-hoc feedback. There is a bar chart above that depicts people's average confidence and accuracy. Below are two bar charts, showing the corresponding situations of ideal confidence and accuracy, as well as the corresponding situations of the current user's confidence and accuracy.}
	\label{fig:condition-interfaces}
\end{figure*}

\textbf{Think the Opposite (Think)}. 
Research suggests that humans' overconfidence in their predictions is a common issue \cite{dunning2003people, kahneman2011thinking}. This often occurs due to biases like anchoring \cite{nourani2021anchoring, furnham2011literature, rastogi2020deciding} and confirmation bias \cite{nickerson1998confirmation}. People tend to favor information that supports their views, making it challenging to consider alternatives in decision-making \cite{considine2012thinking}. To improve self-confidence calibration, we design an intervention inspired the ``pre-mortem'' proposed by Klein \cite{klein2007performing} and Mitchell \cite{mitchell1989back}. Participants are asked to imagine a scenario where their initial decision was actually wrong, encouraging them to think beyond their initial perspective \cite{kahneman2011thinking}. Based on this theory, we introduce ``Thinking the Opposite'' (Figure \ref{fig:condition-interfaces} (a)), where users, before reporting their self-confidence, need to respond to two questions: (1) ``Which features of this profile might favor an alternative prediction?'' and (2) ``If your prediction is incorrect, what could be the most likely reason for that?''. By answering these two questions, users are expected to quantify their confidence more carefully.

\textbf{Thinking in Bets (Bet)} leverages insights from works of Moore \cite{moore2020perfectly} and Duke \cite{duke2019thinking}, who proposed to adjust human self-confidence by using ``betting'' to incentivize careful consideration of one's confidence level. In our design (Figure \ref{fig:condition-interfaces} (b)), participants receive a 200-coin bonus account. They are prompted to decide whether and how much they want to bet on their predictions for each task (i.e., we ask them ``Want to bet? How many coins do you want to bet on your prediction?''), with their account balance adjusted based on prediction accuracy and the amount bet. For instance, a correct prediction with a 10-coin bet results in a 10-coin addition to their account, while an incorrect prediction with a 3-coin bet deducts 3 coins. Participants are informed that their coins will be converted to bonuses at a rate of 200 points to \$1 after they finish the experiment. Note that real-time balance updates are not provided to prevent participants from knowing the ground truth.

\textbf{Calibration Status Feedback (Feedback)} aligns with Moore's recommendation \cite{moore2020perfectly} to provide evidence-based assessments of performance or probability for self-confidence calibration. To achieve this in decision-making, we introduce ``Calibration Status Feedback'', offering two feedback interfaces. The first is a real-time feedback interface (Figure \ref{fig:condition-interfaces} (c, left)), providing immediate feedback after each prediction. Users receive information about the actual answer and their self-confidence status, categorized as match, over-confident, or under-confident. A historical confidence status is visually represented as a colored block and continually updated in the status bar. The second interface is an overall post-hoc feedback interface (Figure \ref{fig:condition-interfaces} (c, right)), offering both an overview (Figure \ref{fig:condition-interfaces} (1)) and a detailed analysis (Figure \ref{fig:condition-interfaces} (2)). The overview summarizes the proportions of \emph{match}, \emph{over-confident}, and \emph{under-confident} instances from past feedback sessions. It also calculates the user's accuracy and average confidence, providing a high-level summary like ``Based on your past predictions, you tend to be over-confident''. In the detailed analysis section, the user's historical predictions are segmented into five bins based on confidence distribution. For each bin, accuracy and average confidence are computed and visualized as a reliability diagram \cite{brocker2007increasing}. An ``ideal'' reliability diagram is presented for reference, depicting accurate alignment between confidence and accuracy to help users discern the disparity between their self-confidence and ``appropriate'' self-confidence levels. Previous research investigating human confidence has studied real-time and post-hoc feedback separately \cite{sharp1988performance, perfect2000practice, gonzalez2007aligning}. We combined these two feedback types for two reasons. First, using only real-time feedback might limit participants to remembering their most recent confidence levels, making it hard for them to have a comprehensive understanding and recall of their confidence status across the entire 20 task instances. Second, relying solely on overall feedback could obscure which instances lead to over- or under-confidence. While this combination may not be perfect, we encourage further exploration of more effective feedback designs.

\subsection{Conditions}
In this between-subjects study, to minimize potential interference, all participants are tasked with making predictions without AI assistance. Participants are randomly assigned to one of four conditions:

\begin{itemize}
    \item \textbf{Think the Opposite (Think)}: In the main task, participants made their decisions with the \emph{Think the Opposite} interface (Figure \ref{fig:condition-interfaces} (a)). Before indicating their confidence, participants had to think of features/attributes that might make the actual answer contrary to their initial prediction and give their reasons.
    \item \textbf{Thinking in Bets (Bet)}: Using the \emph{Thinking in Bets} interface (Figure \ref{fig:condition-interfaces} (b)) in the main task, participants were invited to bet on their predictions (0-10 coins) before indicating their confidence.
    \item \textbf{Calibration Status Feedback (Feedback)}: Participants first engaged in a feedback session with the \emph{Calibration Status Feedback} interface (Figure \ref{fig:condition-interfaces} (c)), and subsequently move on to the main task (without feedback anymore).
    \item \textbf{Control}: No calibration is applied in the main task. Participants just made their decisions and indicated their confidence.
\end{itemize}

\subsection{Research Questions}
Focusing on the main research question \textbf{RQ2: How can humans' self-confidence be calibrated and how will different self-confidence calibration mechanisms affect humans' perceptions and user experience?}, we raise two sub-questions:
\begin{itemize}
    \item[-] \textbf{RQ 2.1}: How will different self-confidence calibration mechanisms affect humans' task performance and the appropriateness of their self-confidence?
    \item[-] \textbf{RQ 2.2}: How will different self-confidence calibration mechanisms affect humans' perceptions (e.g., perceived self-confidence appropriateness, performance, and complexity) and user experience (e.g., mental demand, preference, and satisfaction)?
\end{itemize}



\subsection{Task and AI Model}
In this study, we continue to employ income prediction as our decision task, similar to Study 1. We selected 10+20 task instances, following the data selection criteria established in Study 1. Each participant, in every condition, first provides 10 predictions without calibration and then proceeds to make 20 calibrated predictions. Notably, the \emph{Feedback} condition includes an extra feedback session, requiring participants to make 20 additional predictions.

\subsection{Procedure}
Participants followed this experimental process:
\begin{enumerate}[leftmargin=0cm, itemindent=0.5cm]
    \item \textbf{Tutorial}: Upon consenting, participants were given a tutorial on the meanings and value ranges of attributes in the profile table, including the income distribution per attribute from the training dataset. Understanding was verified via qualification questions, allowing only those with correct answers to proceed.

    \item \textbf{Familiarization task}: Participants then completed the first 10 tasks to familiarize themselves with the task nature, without ground truth information or calibration.

    \item \textbf{Calibration Mechanism Tutorial}: Participants learned about their assigned calibration mechanism. In the \emph{Think} and \emph{Bet} conditions, they experimented with a calibration interface. In the \emph{Feedback} condition, they participated in a feedback session. The \emph{Control} condition skipped this step.
    
    \item \textbf{Main Task}: Moving to the main task (20 cases), participants expressed their confidence using the experimental interface, calibration included or not. The session incorporated two attention checks to ensure data quality.
    
    \item \textbf{Exit Survey}: Participants concluded with a survey, providing feedback on their experience.
\end{enumerate}

\subsection{Participants}
Before recruiting participants, we calculated the required sample size via a power analysis for the four groups using G*Power \cite{faul2009statistical}. We set the default effect size $f$ = 0.25 (indicating a moderate effect), a significance threshold $\alpha$ = 0.05, and a statistical power ($1 - \beta$) = 0.9. This yielded a necessary sample size of 232 participants. After obtaining institutional IRB approval, we recruited participants from Prolific\textsuperscript{\ref{prolific}}. After excluding those who failed the attention check, we got 241 valid responses (Think: 57, Bet: 67, Feedback: 55, Control: 62). Among these participants, 117 self-reported as males, 120 as females, and 4 as non-binary. A total of 35 participants were aged 18-29, 74 aged 30-39, 48 aged 40-49, 48 aged 50-59, and 36 aged over 59. Participants also self-rated their knowledge of artificial intelligence: 20 had no knowledge, 176 knew basic AI concepts, 38 had used AI algorithms, and 7 were AI experts. To incentivize high-quality work, participants received a \$1 bonus if their overall accuracy exceeded 90\%. The study lasted approximately 20 minutes, with participants earning an average wage of about \$11 per hour.

\subsection{Evaluation Measures and Analysis}
\subsubsection{Measurements}
In this study, we assess both the appropriateness of human self-confidence and their experience. 

For \emph{the appropriateness of human self-confidence}, as in study 1, we measure \emph{ECE}. We also measure participants' \emph{Over-confident Ratio} and \emph{Under-confident Ratio} to gain a nuanced understanding.

\begin{footnotesize}
$$\textnormal{\textbf{Over-Confident Ratio}} = \frac{\textnormal{Number of incorrect human predictions with high confidence}}{\textnormal{Total number of human initial predictions}},$$

$$\textnormal{\textbf{Under-Confident Ratio}} = \frac{\textnormal{Number of correct human predictions with low confidence}}{\textnormal{Total number of human initial predictions}},$$
\end{footnotesize}

For \emph{user perceptions and user experience}, we employ and adapt the following metrics on a 7-point Likert scale (1: Strongly Disagree, 7: Strongly Agree). Except for the perceived appropriateness of self-confidence, the other questions have been validated by previous AI-assisted decision-making research. We made necessary adaptations to some items based on our specific scenario. For example, we replaced the word ``system'' in the scale used in previous work with ``the decision-making process with the interface'' (because our experimental interface cannot be called a system).

\begin{itemize}[leftmargin=0cm, itemindent=0cm]
    \item Perceived appropriateness of self-confidence: "I think my self-confidence was appropriate (being able to reflect the actual correctness likelihood of my predictions accurately)."
    \item Perceived performance \cite{kocielnik2019will, smith2020no, springer2019progressive}: "I performed well in this income prediction task."
    \item Mental demand \cite{hart2006nasa, ghai2021explainable, buccinca2021trust}: "I found this task mentally demanding."
    \item Perceived complexity \cite{buccinca2021trust}: "The decision-making process with the interface was complex."
    \item Preference \cite{buccinca2021trust, kulesza2012tell, lee2020co, ribeiro2018anchors, yang2020visual}: "I liked the decision-making process with this interface."
    \item Satisfaction \cite{buccinca2021trust, ghai2021explainable}: "I was satisfied with the decision-making process."
\end{itemize}

\subsubsection{Analysis Method}
Based on normality tests, we found most of the collected data did not follow a normal distribution. Therefore, we employed Kruskal–Wallis tests with Bonferroni correction.

\subsection{Results}
\begin{figure*}[htbp]
	\centering 
	\includegraphics[width=0.9\linewidth]{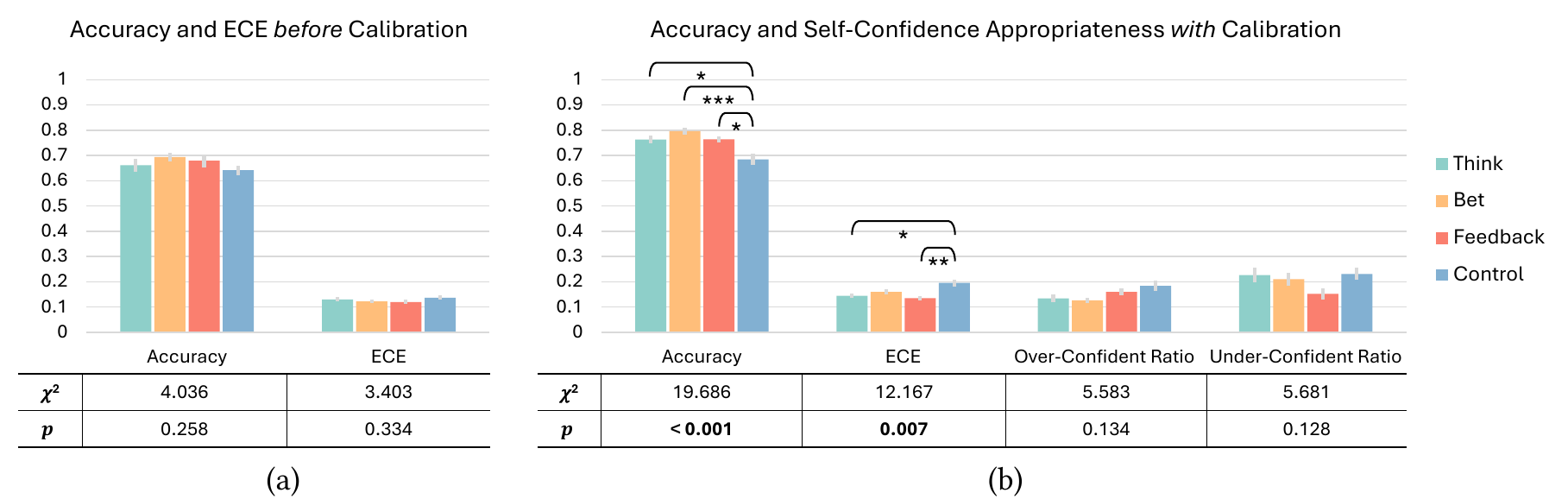}
	\caption{Effects of different calibration conditions on participants' accuracy and the appropriateness of their self-confidence in different conditions. (a) Participants' accuracy and ECE before calibration (in the first 10 familiarization tasks). (b) Participants' accuracy, ECE, Over-Confident Ratio, and Under-Confident Ratio in the main tasks with calibration. Error bars indicate standard errors. (*: $p$ < 0.05; **: $p$ < 0.01; ***: $p$ < 0.001)}
        \Description{This figure has two subgraphs. The left subgraph displays two metrics, and the right subgraph displays four metrics. Each metric contains four bars, from left to right Think, Bet, Feedback, and Control. For the two metrics of the left subgraph, the four bars are almost equally high. In the first metric, accuracy, of the right subgraph, the bars of Think, Bet, and Feedback are all higher than those of Control. In the second metric, ECE, the bars of Think and Feedback are lower than those of Control. In the other two metrics, the four bars are almost equally high.}
	\label{fig:study2_appropriateness}
\end{figure*}

\begin{figure*}[htbp]
	\centering 
	\includegraphics[width=0.9\linewidth]{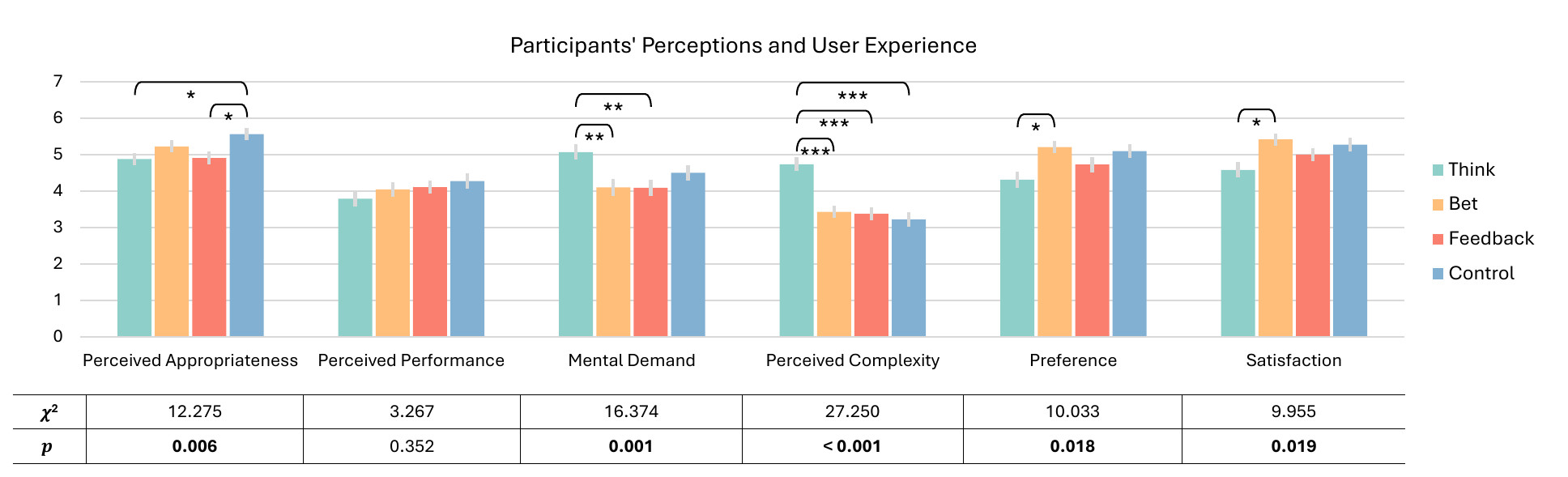}
	\caption{Participants' perceptions and self-reported user experience in different confidence calibration conditions. Error bars indicate standard errors. (*: $p$ < 0.05; **: $p$ < 0.01; ***: $p$ < 0.001)}
         \Description{This figure shows six metrics, and each metric has four bars (from left to right Think, Bet, Feedback, and Control). For the first metric, perceived appropriateness, the bar of Control is higher than the bars of Think and Feedback. For the second metric, perceived performance, the four bars are almost equally high. For the third metric, mental demand, the bar of Think is higher than the bars of Bet, Feedback, and Control. For the fourth metric, perceived complexity, the bar of Think is higher than the other three bars. For the fifth metric, preference, the bar of Bet is higher than the bar of Think. For the last metric, satisfaction, the bar of Bet is higher than the bar of Think.}
	\label{fig:study2_experience}
\end{figure*}

\subsubsection{Effects on Task Performance and the Appropriateness of Self-Confidence (RQ 2.1)}
Figure \ref{fig:study2_appropriateness} shows the effects of different calibration interfaces before calibration (in the first 10 familiarization tasks) and with calibration (in the last 20 main tasks). Since these two batches involved different task samples, we analyzed the participants' accuracy and ECE in the first 10 and last 20 tasks separately. In the first 10 familiarization tasks (Figure \ref{fig:study2_appropriateness} (a)), we found that there was no significant difference in accuracy and ECE between the four conditions before calibration. However, in the main task, with different calibrations, participants' accuracy and ECE were significantly different (Figure \ref{fig:study2_appropriateness} (b)). Specifically, for \textbf{accuracy}, participants in \emph{Think}, \emph{Bet}, and \emph{Feedback} performed significantly better than in \emph{Control} condition. 
This reveals that the calibration mechanism itself can lead to improved task performance compared to no calibration. The reason might be that calibration mobilizes more cognitive resources, makes people think more carefully, and reduces errors caused by insufficient thinking. For \textbf{ECE}, we observed that both \emph{Think} and \emph{Feedback} helped participants maintain a more calibrated self-confidence compared to \emph{Control}. But there was no significant difference between \emph{Bet} and \emph{Control}. This result reveals that although seemingly promising, \emph{Bet} was not effective enough to calibrate participants' self-confidence (perhaps the loss of ``coins'' is not motivating enough). Moreover, we did not find any significant differences among different conditions in terms of \textbf{Over-Confident Ratio} and \textbf{Under-Confident Ratio}. However, we can observe a trend that \emph{Think} and \emph{Bet} might lead to less over-confidence perhaps because participants in these two conditions were guided to think of ``the opposite'' or ``possibility of failure'', which led to more serious quantification of their confidence.


\subsubsection{Effects on Human Perceptions and User Experience (RQ 2.2)}
Figure \ref{fig:study2_experience} shows the effects of different calibration interfaces on participants' perceptions and user experience. We found that participants perceived their self-confidence to be more appropriate in \emph{Control} than in \emph{Think} and \emph{Feedback}, which is very interesting as this subjective result differs from the actual appropriateness of their self-confidence (in \emph{Control}, the appropriateness of human self-confidence is the worst). From this result, we can find that participants have an unreliable perception of their self-confidence without calibration (in \emph{Control}). This also further highlights the necessity to calibrate people's self-confidence.

In addition, participants' mental demands and perceived complexity of the interaction were significantly higher in \emph{Think} than in other conditions. Possible reasons are that we asked people to think about the problem from a second perspective (i.e., the opposite) and asked people to identify features that might lead to opposite results and give reasons for them. This complex process brought more consumption of cognitive resources to the participants, so they felt that the condition was more complex and mentally demanding. Furthermore, we found a trend that \emph{Think} led to lower preference and satisfaction than other conditions. This may be because people are used to adopting heuristics for quick thinking \cite{kahneman2011thinking} and forcing them to analytically think about opposites might change people's usual way of thinking, increase the difficulty of decision-making, and lead to a worse user experience. This finding is consistent with existing work showing that cognitive forcing functions, although making people think more carefully, also degraded the user experience \cite{buccinca2021trust}. Moreover, compared to \emph{Control} condition, \emph{Bet} and \emph{Feedback} did not lead to a worse user experience.

\begin{table*}[hbpt]
\centering  
\fontsize{8}{8}\selectfont  

\caption{Summary of the pros, cons, and applicability of the three calibration mechanisms.}\label{tab:proscons}

\begin{tabular}{m{1.5cm}<{\centering}|m{4cm}<{\centering}|m{4cm}<{\centering}|m{4cm}<{\centering}}
\toprule
\textbf{}&\textbf{Pros}&\textbf{Cons}&\textbf{Applicability}\\
\hline
\textbf{Think Opposite}&Effective for improving the overall appropriateness of human self-confidence; Improving human task performance&Decreasing user experience&Not suitable for ``quick'' decision-making\\
\hline
\textbf{Thinking in Bets}&(Potentially) Effective for reducing over-confidence; Improving human task performance&Ineffective in improving the overall appropriateness of human self-confidence&The incentive ``betting'' mechanism is difficult to apply in certain decision-making tasks\\
\hline
\textbf{Calibration Status Feedback}&Effective for improving the overall appropriateness of human self-confidence; Improving human task performance&Costing more time (due to the feedback session)&Requiring access to some task samples with ground truth for designing the feedback session.\\
\bottomrule
\end{tabular}%
\end{table*}

\subsubsection{A Comprehensive Comparision of Different Calibration Mechanisms}
We also want to analyze the pros, cons, and applicability of our proposed three confidence calibration mechanisms (as shown in Table \ref{tab:proscons}). A general conclusion is that there may not be a perfect calibration mechanism. Specifically, from the aforementioned results, we can see that \emph{Think} is effective for improving the appropriateness of participants' self-confidence but it damages participants' user experience. And since \emph{Think} requires extra investment of cognition resources, it is more suitable for high-stakes tasks and might not be appropriate for ``quick decision-making tasks'' that are time-limited. Besides, although we can observe a trend for \emph{Bet} to reduce over-confidence (not significantly), it is ineffective in improving the overall appropriateness of human self-confidence. And despite that the incentive ``betting'' mechanism is interesting, it can be difficult to apply to certain serious decision-making tasks in which the incentive mechanism cannot be established. What's more, although \emph{Feedback} can effectively improve the appropriateness of human self-confidence without leading to a worse user experience, it can cost more time due to the feedback session. And the feedback session is only feasible when ground truth data is accessible.

Overall, on the one hand, we recommend designers choose a suitable calibration mechanism based on the specific goal and task property. For example, if the only purpose is to reduce humans' over-confidence and the incentive mechanism can be established, \emph{Bet} might be a good choice. On the other hand, designers should not only focus on the effectiveness of confidence calibration but also take the potential negative effects on user experience into consideration. In this paper's task, since it is easy to access the training data with ground truth, considering both the effectiveness of confidence calibration and the harmless effects on user experience, we chose \emph{Feedback} as our calibration mechanism in Study 3 to further explore its effects in AI-assisted decision-making.

\subsubsection{Summary}
This experiment compared the effects of three calibration mechanisms on people's task performance and self-confidence appropriateness, as well as their impacts on users' perceptions and user experience. The experimental results reveal the advantages and disadvantages of different calibration mechanisms and provide designers with insights into the design and selection of calibration strategies. We found that the proposed self-confidence calibration can improve humans' task performance compared to the control condition, but not all calibration mechanisms can improve the appropriateness of human self-confidence (RQ 2.1). Also, \emph{Think} led to participants' worse user experience but \emph{Feedback} and \emph{Bet} did not decrease user experience (RQ 2.2).





%% file: sections/06-Study3.tex
\section{Study 3 - Investigating the Effects of Human Self-Confidence Calibration on AI-Assisted Decision Making}
Our third study aims to explore the effects of the introduction of self-confidence calibration on AI-assisted decision-making.

\subsection{Research Question}
Focusing on our main research question \textbf{RQ3: How will calibration of humans’ self-confidence affect the appropriateness of their reliance on AI’s suggestions and task performance?}, we specifically ask the following sub-questions.

\begin{itemize}
    \item[-] \textbf{RQ 3.1}: How will the calibration of human self-confidence affect humans' reliance behaviors?
    \item[-] \textbf{RQ 3.2}: How will the calibration of human self-confidence affect the appropriateness of human reliance on AI suggestions (e.g., over-reliance, under-reliance)?
    \item[-] \textbf{RQ 3.3}: How will the calibration of human self-confidence affect humans' task performance?
\end{itemize}

\subsection{Task Setup}
The task setup mirrors Study 1 and Study 2, employing income prediction as the decision-making task. 

\subsection{Conditions}
This study delves into the effects of human self-confidence calibration when collaborating with an AI model that displays its confidence, guided by two considerations. First, most contemporary AI models are capable of generating probability estimations, i.e., confidence levels, making it a common and feasible practice. Second, revealing AI confidence is a widely recognized design choice for calibrating human trust in AI-assisted decision-making \cite{zhang2020effect, rastogi2020deciding, zhao2023evaluating, bansal2021does}. Therefore, we focus on scenarios where AI confidence is presented. Under this setting, we explore two conditions:
\begin{itemize}
    \item \textbf{With Calibration}: This condition applies \emph{Calibration Status Feedback} to calibrate participants' self-confidence.
    \item \textbf{No Calibration}: In this condition, we do not apply any confidence calibration mechanisms to participants.
\end{itemize}

\subsection{Procedure}
This between-subjects experiment randomly assigned participants to either the \textbf{Calibration} or \textbf{No Calibration} conditions. Upon entering the experimental interface, participants underwent a tutorial to familiarize themselves with the task, which included qualification questions to ensure they grasped key task knowledge. In the \textbf{Calibration} condition, participants then proceeded to a \textbf{Feedback session} for confidence calibration before entering the main task session. Participants in the \textbf{No Calibration} condition started with demo tasks to become familiar with the task process and then directly engaged in the main task session. During the main task, all participants first made their initial predictions (and indicated their confidence), then saw AI's suggestions with confidence, and finally made their final decisions.

\subsection{Participants}
We first calculated the required sample size via a power analysis for the two groups using G*Power \cite{faul2009statistical}. We set the default effect size $f$ = 0.6 (indicating a moderate effect), a significance threshold $\alpha$ = 0.05, and a statistical power ($1 - \beta$) = 0.8. This yielded a necessary sample size of 90 participants. After obtaining institutional IRB approval, we recruited participants from Prolific\textsuperscript{\ref{prolific}}. Throughout the experiment, we included 2 attention-check questions. After filtering data from inattentive participants, 117 valid responses remained (Calibration: 57, No Calibration: 60). Among these participants, 57 self-reported as males, 55 as females, and 5 as non-binary. Age distribution included 20 participants aged 20-29, 35 aged 30-39, 18 aged 40-49, 26 aged 50-59, and 18 aged over 60. Participants' self-rated knowledge of artificial intelligence varied, with 7 having no knowledge, 76 knowing basic AI concepts, 20 having used AI algorithms, and 14 being AI experts. To incentivize high-quality work, participants received a \$1 bonus in addition to the base payment if their overall accuracy exceeded 90\%. The entire study lasted approximately 20 minutes, with participants earning an average wage of about \$12 per hour.

\subsection{Evaluation Measures and Analysis}
\subsubsection{Measurements}
We measure participants' \emph{self-confidence appropriateness}, \emph{reliance}, \emph{task performance}, and \emph{reliance appropriateness}. For \emph{self-confidence appropriateness}, we use the same metrics as used in Study 1 and 2. 

For \emph{reliance}, we collect the following measures:

\begin{footnotesize}
$$\textnormal{\textbf{Agreement fraction}} = \frac{\textnormal{Number of final decisions same as the AI suggestion}}{\textnormal{Total number of decisions}},$$
$$\textnormal{\textbf{Switch fraction}} = \frac{\textnormal{Number of decisions user switched to agree with the AI model}}{\textnormal{Total number of decisions with initial disagreement}},$$
$$\textnormal{\textbf{Follow high confidence fraction}} = $$
$$\frac{\textnormal{Number of tasks where user followed the prediction with higher condfidence}}{\textnormal{Total number of decisions with initial disagreement}},$$
\end{footnotesize}

    
For \emph{task performance}, we calculate participants' initial accuracy (before seeing AI suggestions) and final accuracy (after seeing AI suggestions).

Additionally, based on our analytical framework, we calculate:

\begin{itemize}[leftmargin=0.1cm, itemindent=0.5cm]
    \item[-] Distribution of different C-C Matching: We calculate the ratio of different human and AI C-C Matching when human initial predictions disagree with AI suggestions:
    
\begin{footnotesize}
$$\frac{\textnormal{Number of predictions in a specific C-C Matching situation}}{\textnormal{Total number of decisions with initial disagreement}},$$
\end{footnotesize}
    
    \item[-] Errors caused by different C-C Matching: We calculate the occurrence of incorrect human final predictions caused by these four types of C-C Matching: 

    \begin{footnotesize}
$$\frac{\textnormal{Number of incorrect final decisions in a specific C-C Matching situation}}{\textnormal{Total number of decisions}},$$
\end{footnotesize}

\end{itemize}


For \emph{reliance appropriateness}, as Study 1, we assess \emph{Under-Reliance} and \emph{Over-Reliance}. Following \cite{he2023knowing}, we also measure \emph{Accuracy-wid} (accuracy with initial disagreement), a stringent metric for evaluating the appropriateness of human reliance on AI. This measure focuses on whether individuals can make correct decisions when initially disagreeing with AI recommendations, thus offering an assessment more independent of human initial accuracy. In contrast, \emph{Under/Over-Reliance} does not account for the initial correctness of human judgments (e.g. an incorrect initial human prediction followed by acceptance of an incorrect AI suggestion is still deemed \emph{over-reliance}). In other words, \emph{Under/Over-Reliance} is more likely to be affected by humans' independent task performance.

\begin{footnotesize}
$$\textnormal{\textbf{Accuracy-wid}} = \frac{\textnormal{Number of correct final decisions with initial disagreement}}{\textnormal{Total number of decisions with initial disagreement}},$$
\end{footnotesize}
\subsubsection{Analysis Method}
For the analysis, since data did not pass the normality test, we used non-parameter tests. Specifically, we compared two unpaired groups via Mann-Whitney U tests and compared two paired groups via Wilcoxon Signed-Rank tests.

\subsection{Results}

\begin{figure}[htbp]
	\centering 
	\includegraphics[width=\linewidth]{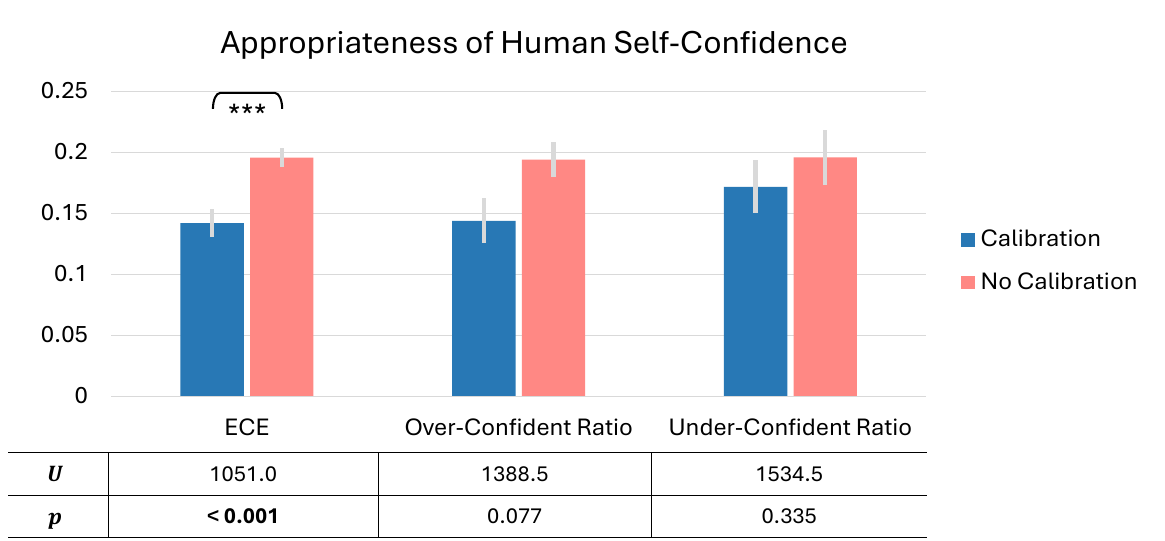}
	\caption{Manipulation check results: the appropriateness of human self-confidence in \textbf{Calibration} and \textbf{No Calibration} conditions. Error bars indicate standard errors. (*: $p$ < 0.05; **: $p$ < 0.01; ***: $p$ < 0.001)}
	\label{fig:study3_selfconfidence}
          \Description{This figure shows three metrics, and each metric has two bars (from left to right Calibration and No Calibration). For the first metric, ECE, the bar of Calibration is lower than the bar of No Calibration. For the second and third metrics, Over-Confident Ratio and Under-Confident Ratio, the bars of Calibration are both lower than the bars of No Calibration but without significance marks.}
\end{figure}
\subsubsection{Manipulation Check} First, we want to verify whether, in the \textbf{Calibration} condition, participants' self-confidence in their initial predictions actually gets calibrated. As shown in Figure \ref{fig:study3_selfconfidence}, Mann-Whitney U tests reveal that in the \textbf{Calibration} condition, participants' \emph{ECE} score is significantly lower than that in the \textbf{No Calibration} condition. It reveals that participants' overall accuracy and confidence are better matched given calibration. Although when analyzed separately, no significant improvements are found in \emph{Over-Confident Ratio} and \emph{Under-Confident Ratio}, notable decrease trends can be observed. These results confirm the effectiveness of our manipulation.

\begin{figure}[htbp]
	\centering 
	\includegraphics[width=\linewidth]{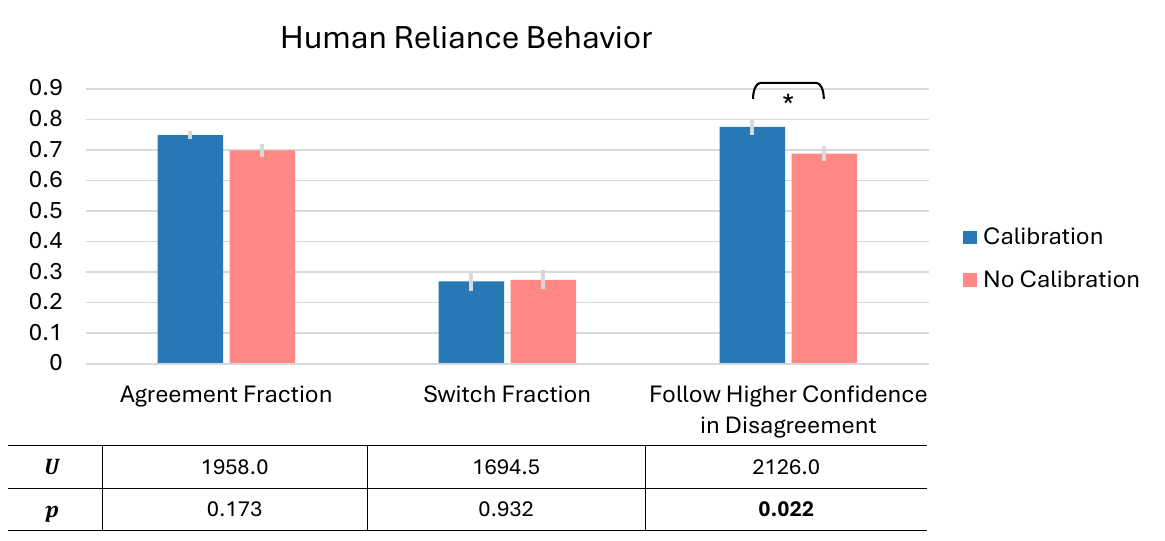}
	\caption{Human reliance behaviors in \textbf{Calibration} and \textbf{No Calibration} conditions. Error bars indicate standard errors. (*: $p$ < 0.05; **: $p$ < 0.01; ***: $p$ < 0.001)}
       \Description{This figure shows three metrics, and each metric has two bars (from left to right Calibration and No Calibration). For the first and second metrics, Agreement Fraction and Switch Fraction, the bar of Calibration is similar high as the bar of No Calibration. For the third metric, Follow High Confidence in Disagreement, the bar of Calibration is higher than the bar of No Calibration with a significance mark.}
	\label{fig:reliance}
\end{figure}

\subsubsection{Effects on Human Reliance Behaviors (RQ 3.1)}
\label{sec_followhigh}
The calibration of human self-confidence did not lead to significant differences in terms of \emph{Agreement Fraction} and \emph{Switch Fraction} (see Figure \ref{fig:reliance}). However, we observed that when their initial predictions differed from AI's suggestions, participants in the \textbf{Calibration} condition more often followed the member (human or AI) who had higher confidence compared to those in the \textbf{No Calibration} condition. Since higher confidence can reflect a higher correctness likelihood especially when both human and AI's confidence is calibrated, this result suggests that the calibration of people's self-confidence promotes a more rational utilization of the confidence information.

\begin{figure}[htbp]
	\centering 
	\includegraphics[width=\linewidth]{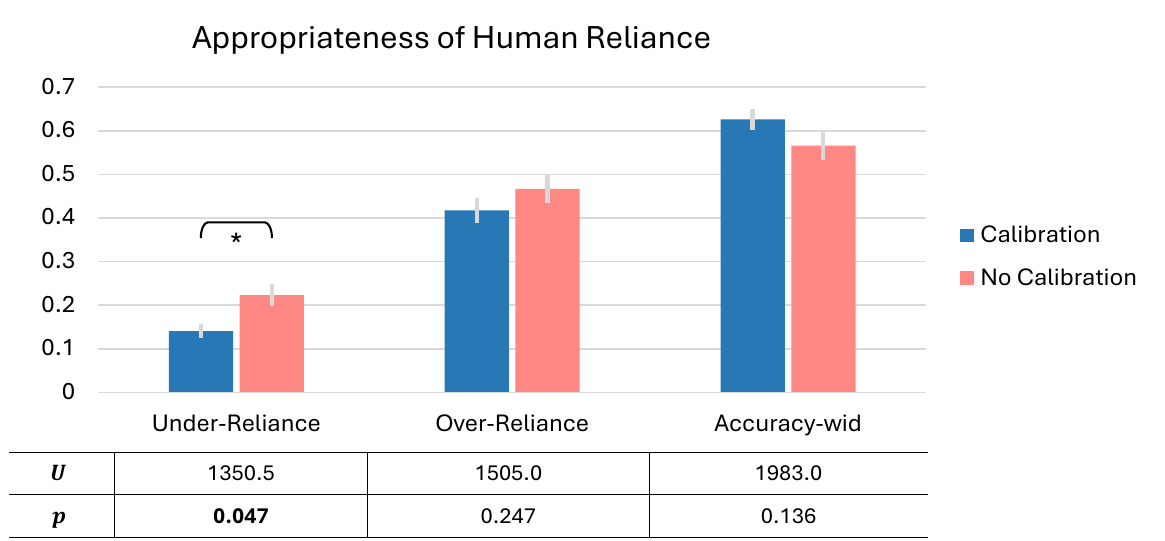}
	\caption{The appropriateness of human reliance. Error bars indicate standard errors. (*: $p$ < 0.05; **: $p$ < 0.01; ***: $p$ < 0.001)}
	\label{fig:appropriatereliance}
        \Description{This figure shows three metrics, and each metric has two bars (from left to right Calibration and No Calibration). For the first metric, Under-Reliance, the bar of Calibration is lower than the bar of No Calibration with a significance mark. For the second and third metrics, Over-Reliance and Accuracy-wid, the two bars are similarly high.}
\end{figure}

\subsubsection{Effects on the Appropriateness of Human Reliance (RQ 3.2)}

Results reveal that participants in the \textbf{Calibration} condition exhibited significantly lower \emph{Under-Reliance} than those in the \textbf{No Calibration} condition (Figure \ref{fig:appropriatereliance}). However, no significant difference is observed in terms of \emph{Over-Reliance} and \emph{Accuracy-wid}. This means that calibrating people's self-confidence only improves the appropriateness of people's reliance in some aspects but not all. We will further analyze the possible reasons in the Discussion (Sec \ref{sec:complicated}).

\begin{figure}[htbp]
	\centering 
	\includegraphics[width=0.85\linewidth]{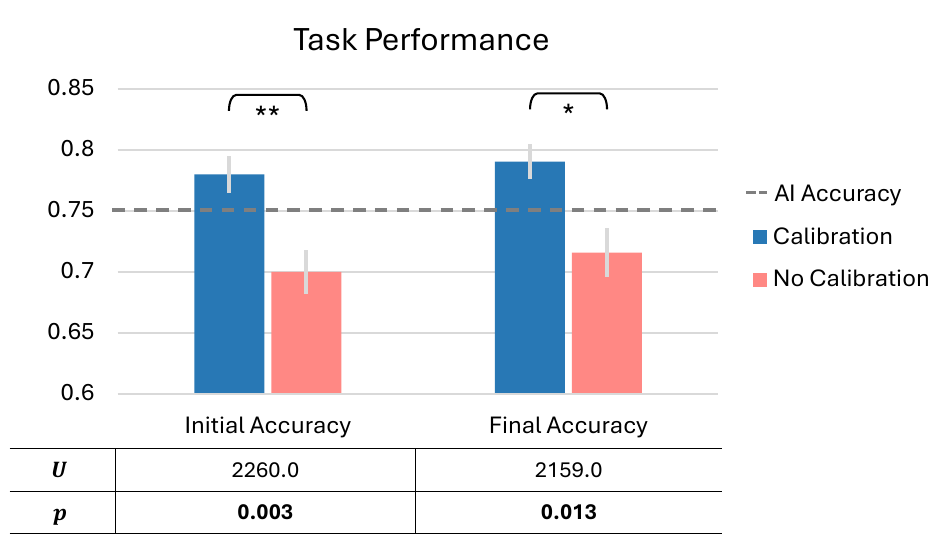}
	\caption{Humans' initial and final performance. Error bars indicate standard errors. (*: $p$ < 0.05; **: $p$ < 0.01; ***: $p$ < 0.001)}
         \Description{This figure shows two metrics, and each metric has two bars (from left to right Calibration and No Calibration). For the first metric, Initial Accuracy, the bar of Calibration is higher than the bar of No Calibration with a significance mark. For the second metric, Final Accuracy, the bar of Calibration is also higher than the bar of No Calibration. The bar of Calibration in the second metric is a little higher than that in the first metric. The bar of No Calibration in the second metric is also a little higher than that in the first metric.}
	\label{fig:performance}
\end{figure}

\begin{figure*}[htbp]
	\centering 
	\includegraphics[width=\linewidth]{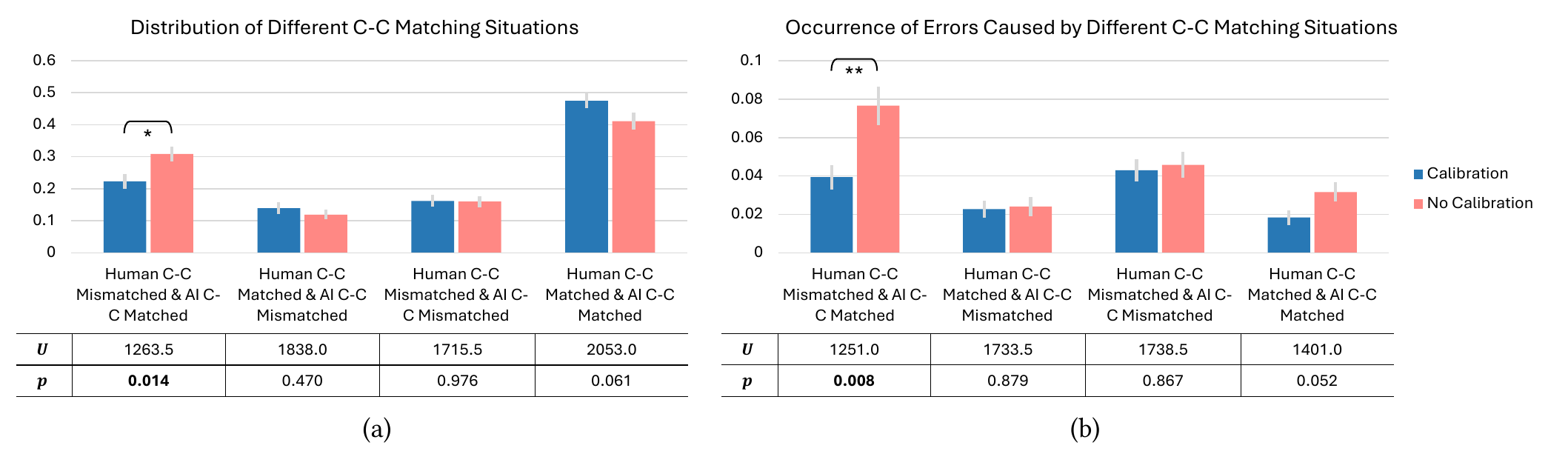}
	\caption{A detailed analysis comparing \textbf{Calibration} and \textbf{No Calibration} based on the proposed analytical framework. (a) The distribution of different human-AI C-C Matching situations. (b) The occurrence of error caused by different human-AI C-C Matching situations. Error bars indicate standard errors. (*: $p$ < 0.05; **: $p$ < 0.01; ***: $p$ < 0.001)}
	\label{fig:study3_distribution}
          \Description{This figure shows two subgraphs. The subgraph on the left represents the distribution of different C-C Matching situations, which includes four metrics. The subgraph on the right represents the occurrence of errors caused by different C-C Matching situations, which includes four metrics. Each metric has two bars, namely Calibration and No Calibration. In both subgraphs, only in the first metric, Human C-C Mismatched\&AI C-C Matched, the bar for Calibration is lower than that for No Calibration, with a significance mark. In the other three metrics, both bars are similarly high.}
\end{figure*}

\begin{figure}[htbp]
	\centering 
	\includegraphics[width=0.85\linewidth]{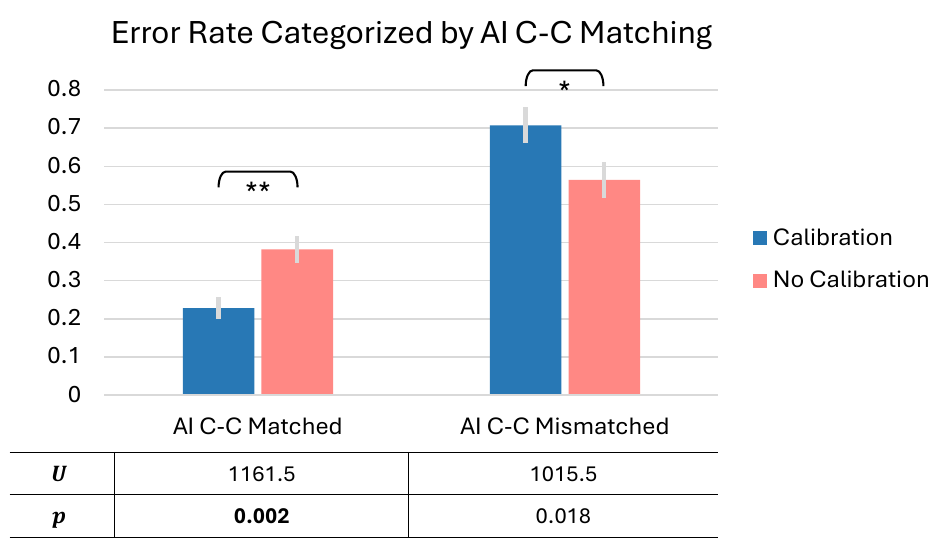}
	\caption{A detailed analysis of error rate when AI's confidence in a prediction matches the prediction's correctness (\emph{AI C-C Matched}) and when AI's confidence in a prediction mismatches the prediction's correctness (\emph{AI C-C Mismatched}). Error bars indicate standard errors. (*: $p$ < 0.05; **: $p$ < 0.01; ***: $p$ < 0.001)}
	\label{fig:detailedanalysis}
           \Description{This figure shows two metrics, the first being AI C-C Matched and the second being AI C-C Mismatched. Each metric has two bars, namely Calibration and No Calibration. In the first metric, the bar for Calibration is lower than the bar for No Calibration with a significance mark. In the second metric, the bar for Calibration is higher than the bar for No Calibration with a significance mark.}
\end{figure}

\subsubsection{Effects on Task Performance (RQ 3.3)}

Figure \ref{fig:performance} presents participants' task performance (their initial accuracy and final accuracy in the two conditions). Mann-Whitney U tests showed that both participants' initial and final accuracy in the \textbf{Calibration} condition surpassed those in the \textbf{No Calibration} condition.

Notably, in the \textbf{Calibration} condition, participants' final accuracy outperformed both the accuracy of AI alone (0.75) and their initial accuracy. In contrast, in the \textbf{No Calibration} condition, neither participants' initial or final accuracy outperformed AI alone. It indicates the potential of self-confidence calibration for achieving complementary team performance \cite{bansal2021does, schemmer2023appropriate}. 
We note that Wilcoxon Signed-Rank tests did not show statistically significant differences between participants' initial and final performance in the \textbf{Calibration} condition. However, the non-significance does not mean that participants did not pay attention to AI suggestions. From the participants’ \emph{Switch Fraction} (the fraction of cases where participants changed their initial predictions after seeing AI’s recommendation) in Figure \ref{fig:reliance}, we can see that participants changed their views 27\% of the time when they disagreed with the AI's views. This indicates that participants’ decisions were influenced by AI’s recommendation. The reason behind the non-significance in accuracy improvement may be that self-confidence calibration also improves their initial performance (similar findings can be seen in Study 2). \emph{The improved initial performance and reduced inappropriate reliance on AI might jointly act on participants' improved final performance.} Since the participants' initial performance ($M=0.78, SD=0.11$) was already higher than AI (0.75), it was difficult to achieve further significant improvements in final performance after working with an AI assistant with slightly lower accuracy. 

\subsubsection{Effects on the Distribution of Different Disagreements and Errors}

Based on the proposed analytical framework, we further dig into task performance. Figure \ref{fig:study3_distribution} (a) displays the distribution of different C-C Matching situations when human-AI disagreements occurred. In the \textbf{Calibration} condition, the \emph{Human C-C Mismatched \& AI C-C Matched} situation is significantly lower than in the \textbf{No Calibration} condition. Additionally, there is a trend indicating a higher \emph{Human C-C Matched \& AI C-C Matched} in the \textbf{Calibration} condition compared to that in the \textbf{No Calibration} condition. These findings suggest that calibrating human self-confidence aligns human confidence more closely with their actual correctness, reducing mismatches occurring from the human side.

Figure \ref{fig:study3_distribution} (b) shows a detailed analysis of the occurrence of errors caused by different C-C Matching situations. We can see a significant reduction in errors caused by \emph{Human C-C Mismatched \& AI C-C Matched} in the \textbf{Calibration} condition compared to the \textbf{No Calibration} condition. This indicates that while AI-side errors remain uncontrollable, human self-confidence calibration effectively reduces errors originating from the human side.

We further categorize participants' decisions only based on the C-C Matching of the AI side (Figure \ref{fig:detailedanalysis}). We find that in task cases where \emph{AI C-C Matched}, \textbf{Calibration} results in significantly fewer error rates compared to \textbf{No Calibration}. Conversely, in task cases where \emph{AI C-C Mismatched}, \textbf{Calibration} leads to significantly more errors than \textbf{No Calibration}. The reason can be tied back to the results in participants' reliance behavior (Sec. \ref{sec_followhigh}), calibrating self-confidence makes participants act more rationally (relying more on the one who holds higher confidence). When \emph{AI C-C Matched}, the AI gives correct recommendations with high confidence or incorrect recommendations with low confidence, following high confidence will lead to a higher likelihood to be correct. On the contrary, when \emph{AI C-C Mismatched}, the AI gives correct recommendations with low confidence or incorrect recommendations with high confidence. At this time, following high confidence will lead to more errors. However, it's important to recognize that when an AI's confidence is accurately calibrated, instances of \emph{AI C-C Matched} will significantly outnumber those of \emph{AI C-C Mismatched}. Therefore, calibrating human self-confidence accordingly should result in more benefits than drawbacks.





\subsubsection{Summary}
In general, calibrating people's self-confidence makes people act more rationally when only confidence information exists (RQ 3.1). However, confidence calibration only reduced under-reliance in our setting (RQ 3.2). Furthermore, self-confidence calibration improves people's initial accuracy and final accuracy (RQ 3.3). A detailed analysis shows that the performance improvement may be due to the reduced occurrence of \emph{C-C Mismatched} and the corresponding errors on the human side. Additionally, we found when \emph{AI C-C Matched}, the calibration of human self-confidence significantly reduced the error rate. We also discovered when the confidence of the AI does not match its correctness, the calibration of human confidence has a negative effect. Therefore, more future work is needed to specifically address this issue.

%% file: sections/07-Discussion.tex
\section{Discussion}
This paper underscores the pivotal role of the appropriateness of human self-confidence in AI-assisted decision-making. Our proposed analytical framework integrates the confidence-correctness matching from both human and AI perspectives. Through this framework, we delve into the causes of humans' inappropriate reliance and analyze the potential of calibrating human self-confidence.

Our research comprises three consecutive empirical studies centered on understanding the effects of self-confidence calibration in AI-assisted decision-making. The first study investigates the relationship between human self-confidence appropriateness and the appropriateness of their reliance, emphasizing the significance of improving human self-confidence appropriateness in decision-making. Building upon cognitive science theories, the second study proposes three calibration mechanisms, empirically assessing their impact on self-confidence appropriateness and user experience. Following this, the third study incorporates the calibration of human self-confidence into the AI-assisted decision-making process, uncovering its advantages as well as its limitations. In this section, we delve into a comprehensive discussion of our principal findings and offer insights into design implications.

\subsection{Inappropriate Reliance: Over/Under-Trust in AI OR Under/Over-Confident in Oneself?}

Previous studies on AI-assisted decision-making have often focused on calibrating people's trust in AI when promoting appropriate reliance \cite{zhang2020effect, bansal2021does, zhao2023evaluating}. They attribute that over-reliance on AI stems from over-trusting AI, while under-reliance on AI results from under-trusting AI. However, we argue that this attribution oversimplifies the issue without adequately considering the confidence factor from the human perspective.

We propose that inappropriate reliance can arise from both inappropriate trust in AI and inappropriate confidence in oneself. For instance, when people adopt an incorrect suggestion from AI, it might be due to both over-trusting AI and being under-confident. The reasons for these two behaviors differ significantly. Over-trusting AI might result from automation bias \cite{cummings2004automation} or be influenced by certain elements/information of AI, such as stated accuracy \cite{yin2019understanding}, high confidence levels \cite{zhang2020effect}, or seemingly convinced explanations \cite{bansal2021does}. On the other hand, under-confidence in oneself might stem from insufficient task expertise, recent task failures, incomplete task information, or other psychological factors \cite{moore2017confidence}.
By conducting retrospective analyses with our proposed analytical framework, designers can identify root causes and make targeted improvements to specific aspects of the human-AI system. Our proposed analytical framework, which examines inappropriate reliance from two perspectives, can complement existing research, offering a new viewpoint to comprehend the basis of inappropriate reliance.


\subsection{Rational Reliance with Limited Information}

In AI-assisted decision-making, where ground truth remains unknown to both humans and AI, it's crucial to make nuanced reliance decisions based on calibrated confidence \cite{guo2017calibration, zhang2020effect}. When faced with disagreements and access to confidence levels from both parties, adopting the suggestion of the higher-confidence party is a rational choice, as our proposed calibration conditions suggest.
However, it's essential to acknowledge that this rational behavior doesn't consistently lead to more accurate decisions. Study 3, for instance, yielded mixed outcomes. While calibrating human self-confidence reduced error rates significantly when \emph{AI C-C Matched}, error rates increased substantially in cases of \emph{AI C-C Mismatched}. Unfortunately, \emph{AI C-C Mismatched} cannot be eliminated even when AI's confidence is well-calibrated. Thus, addressing people's reliance in \emph{AI C-C Mismatch} necessitates an approach beyond human self-confidence calibration, potentially involving educating individuals about AI error boundaries to make more nuanced judgments \cite{bansal2019beyond}.

\subsection{The Complicated Relationship between the Appropriateness of Self-Confidence and the Appropriateness of Reliance}
\label{sec:complicated}

In Study 3, we found that calibrating human self-confidence reduces under-reliance but doesn't significantly impact over-reliance and accuracy-wid. We posit three possible reasons for such insufficient improvement. First, while calibrating human self-confidence improves its appropriateness, the extent of improvement falls short. This is evident in Figure \ref{fig:study3_selfconfidence}, where the \emph{Over-Confident Ratio} and \emph{Under-Confident Ratio} in the \textbf{Calibration} condition only show trends towards decrease but lack statistical significance. Therefore, future work needs to design more effective self-confidence calibration mechanisms to further enhance the appropriateness of people's self-confidence. Second, self-confidence calibration also brings side effects, that is, when \emph{AI C-C Mismatched}, because people tend to follow the prediction of the party with higher confidence, the error rate increases compared with no calibration. Third, the relationship between self-confidence and reliance on AI is intricate and nonlinear. This complexity aligns with findings in He et al.'s study \cite{he2023knowing}. Inappropriate self-confidence is only one of many causes of inappropriate reliance. 
Despite there being a significant correlation between the appropriateness of self-confidence and the appropriateness of reliance (Study 1), note that correlation does not equal causation. 
We envision a more complex interplay of factors influencing this relationship, including humans' expertise, AI literacy, intrinsic trust in AI, and cognitive biases, among others. Thus, \emph{solely calibrating self-confidence may not suffice to foster appropriate reliance on AI}. Future research is needed for a deeper understanding of this intricate logic.


\subsection{Multifaceted Effects of Self-Confidence Calibration on Final Task Performance} We observed an improvement in humans' task performance with self-confidence calibration. Based on our analysis, the improved task performance results from three pivotal factors. The first factor is improved human independent accuracy. Study 2 and Study 3 demonstrate that in the \textbf{Calibration} condition, individuals exhibit significantly improved initial performance compared to the \textbf{No Calibration} condition. This enhancement might arise from increased engagement in System 2 thinking during the calibration process, reducing errors resulting from inadequate analytical thinking \cite{kahneman2011thinking}. The second factor is improved self-confidence appropriateness. Self-confidence calibration directly reduces the occurrence of \emph{Human C-C Mismatch} and diminishes errors stemming from such mismatch. The third factor is humans' appropriate reliance on AI. Calibration helps people make more rational reliance choices when facing disagreements.

In summary, the final task performance improvement is a product of multifaceted factors. This insight suggests that to enhance human-AI collaboration, apart from focusing on improving AI performance, designers should treat the human-AI collaboration as a whole system, making efforts from diverse perspectives.

\subsection{Implications for Future AI-Assisted Decision-Making Interface Design} Based on the key findings from our studies, we present several design recommendations for designers' consideration.

\textbf{DR1: Calibrating User Self-Confidence Before Initiating Tasks.}
The results from Study 1 suggest that inappropriate user self-confidence exacerbates inappropriate reliance. Therefore, before designing or deploying AI-assisted decision-making systems, it's crucial to gather users' prediction data through a testing phase to understand users' self-perceived competence (i.e., confidence) in the current task. If users' self-confidence is unreliable, interventions to calibrate user confidence should be implemented in such cases. Alternatively, designers may consider making confidence calibration a default setup in AI-assisted decision-making.

\textbf{DR2: Diagnosing and Improving the System Using the Analytical Framework.}
We encourage designers, during the iterative phase of AI-assisted decision-making system development, to use our proposed analytical framework to ``\emph{diagnose}'' the confidence calibration status of users and AI, analyze the underlying causes of inappropriate reliance, and make targeted improvements. For instance, developers or designers can statistically analyze user prediction data from both the Human and AI perspectives regarding C-C Matching. If a significant C-C Mismatch exists among users, designers should introduce calibration mechanisms to refine users' self-confidence. Conversely, if AI C-C Matching issues prevail, designers should collaborate with AI algorithm engineers to optimize confidence calibration in AI.

\textbf{DR3: Choosing Suitable Calibration Based on Calibration Purpose and Task Properties.}
There's no one-size-fits-all calibration mechanism. When choosing calibration methods, designers should compare various options based on the calibration's purpose and task properties. For instance, if target users tend to be overconfident in the selected task, reducing their confidence can be achieved by emphasizing the cost of decision errors (e.g., \emph{Bet}) or encouraging them to approach problems from the opposite perspective (e.g., \emph{Think}). If the task itself provides ground truth data for training, designing a feedback session to calibrate user confidence could be effective.

\textbf{DR4: Considering User Experience in Calibration Design.}
When designing calibration interfaces, designers should not only test calibration effectiveness but also consider its impact on user experience. Some calibration methods might demand higher cognitive effort from users, leading to increased cognitive burden and decreased satisfaction. Therefore, these methods might not suit certain user groups averse to critical thinking. Designers should incorporate user experience and tailor calibration methods according to the attributes of target users. For instance, designers can pre-assess target users' intrinsic cognitive motivation through Need for Cognition (NFC) scales \cite{cacioppo1984efficient} and then design interventions accordingly. For example, employing \emph{Think} on people with high NFC.

\textbf{DR5: Using Self-Confidence Calibration for Training Purposes.}
Our results revealed that calibrating user confidence directly enhances users' independent task performance (see Study 2 and 3). Although calibration differs from formal training, it effectively fulfills a similar role. Consequently, designers have the opportunity to utilize confidence calibration as a training method or to enhance user task performance further by integrating calibration with traditional training approaches.

\textbf{DR6: Ensuring AI Confidence Calibration Before Human Confidence Calibration.}
Results from Study 3 indicate that calibrating human confidence only significantly enhances reliance appropriateness when \emph{AI C-C Matched}. Calibrating human confidence in the case of \emph{AI C-C Mismatched} may lead to more incorrect reliance. Therefore, if designers aim to incorporate human self-confidence calibration into decision-making interfaces, it is essential to assess the degree of AI confidence calibration beforehand.

\textbf{DR7: Guiding Rational Use of Confidence Information.}
In situations where both human and AI confidence are well calibrated, following the viewpoint of the higher-confidence party may yield higher expected benefits. While our results indicate an increase in the ``follow high confidence fraction'', there's still room for improvement. Designers can implement additional designs to directly enhance people's comprehension of the concepts and benefits of calibrated confidence and encourage people to consider the predictions of the party with higher confidence more seriously.

\subsection{Limitations and Future Work}

\subsubsection{Extending to multi-level confidence} In our analytical framework, we categorize human and AI's confidence into two levels, which is not fine-grained enough. Actually, our analytical framework can be generalized to multi-level confidence. However, this will bring combinatorial complexity as we need to not only consider whether the confidence-correctness is matched, but also consider the degree of matching (e.g., [correct \& 95\% confidence] is more matched than [correct \& 85\% confidence]). Moreover, the level of confidence can also be the relative value between a human and an AI. For example, if AI confidence is 55\%, then a person's 65\% confidence can be seen as ``higher'' confidence. At the current stage, the two-level division is enough to give us preliminary insights. In future work, we will expand the proposed analysis framework to more fine-grained confidence, and explore whether it can be used to analyze the relative confidence levels of humans and AI.

\subsubsection{Generalizability of Proposed Calibration Mechanisms}
The analysis of the appropriateness of human self-confidence requires individuals to make independent judgments before accessing AI suggestions, which may not suit scenarios prioritizing efficiency and one-stage decision-making. In addition, gathering users' confidence requires users' extra effort (although slight) and may not mirror natural decision-making processes where users form implicit confidence in their minds. Future research should explore methods to infer confidence accurately from decision-making behaviors, like users' hesitation. Additionally, the three self-confidence calibration mechanisms we proposed all have limitations: \emph{Feedback} relies on historical data for calibration; \emph{Think Opposite} may impose cognitive burdens; and \emph{Bet} isn't universally applicable due to its reliance on economic incentives. Nonetheless, the concept of human self-confidence calibration and the analytical framework can be applied to diverse decision-making tasks. Future work is encouraged to devise more effective self-confidence calibration strategies for specific decision-making scenarios.

\subsubsection{Personalized Calibration}
Different calibration strategies offer unique advantages. For example, \emph{Think} reduces over-confidence and may benefit those prone to overconfidence. Similarly, \emph{Bet} may affect different groups differently; it could boost the confidence of risk-takers while reducing the confidence of risk-averse individuals. This paper only explores the effectiveness of diverse calibration strategies on the whole population. A potential future direction is to investigate user-personalized calibration, tailoring methods to individuals with different characteristics.

\subsubsection{Calibrating Humans' Perceptions of AI Confidence}
One interesting finding is that even when humans and AI are both C-C Matched, humans can still make incorrect decisions. Our paper only calibrates humans' self-confidence. Humans' perception of AI's confidence may also need to be calibrated. Future research should explore ``dual calibration'', investigating the combined impact on AI-assisted decision-making.


\section{Conclusion}
This paper provides a comprehensive understanding of the impact of human confidence calibration in AI-assisted decision-making. We first propose a novel analytical framework to parse inappropriate reliance from the perspective of human and AI confidence-correctness matching. Through three user studies, we make three contributions: (1) analyzing the relationship between the appropriateness of human self-confidence and the appropriateness of human reliance, (2) designing and comparing different confidence calibration mechanisms, and (3) examining the impact of human confidence calibration on humans' behavior, task performance, and reliance appropriateness when working with AI that shows confidence. In summary, this paper provides a unique perspective on understanding and promoting humans' appropriate reliance in AI-assisted decision-making, shedding light on the calibration of human self-confidence. We hope that our research will enrich the understanding and discourse within the research community on this topic and pave the way for further research on human self-confidence calibration in human-AI collaboration.